\documentclass[physrev,nofootinbib,reprint]{revtex4-2}

\usepackage{amsmath,amssymb}

\usepackage[final]{graphicx}
\usepackage{relsize} 
\usepackage{bm}
\usepackage{overpic}
\usepackage[usenames,dvipsnames]{xcolor}

\usepackage{siunitx}
\ifdefined\qty\else
  \ifdefined\NewCommandCopy
    \NewCommandCopy\qty\SI
  \else
    \NewDocumentCommand\qty{O{}mm}{\SI[#1]{#2}{#3}}
  \fi
\fi
\ifdefined\unit\else
  \ifdefined\NewCommandCopy
    \NewCommandCopy\unit\si
  \else
    \NewDocumentCommand\unit{O{}m}{\si[#1]{#2}}
  \fi
\fi

\usepackage{natbib}

\usepackage{microtype}

\usepackage[colorlinks,citecolor=red,urlcolor=blue]{hyperref}
\IfFileExists{hypernat.sty}{\usepackage{hypernat}}

\newcommand{\defn}[1]{\emph{#1}}

\newcommand{\bivec}[1]{\tensor{#1}} 

\newcommand{\cwbivec}{{\setlength{\fboxsep}{1pt}\fbox{$\circlearrowright$}}}
\newcommand{\ccwbivec}{{\setlength{\fboxsep}{1pt}\fbox{$\circlearrowleft$}}}

\newcommand{\fourvec}[1]{\bm{\mathsf{#1}}}

\DeclareMathOperator{\tr}{tr}
\DeclareMathOperator{\diag}{diag}

\newcommand{\dblcdot}{\!:\!}

\newcommand{\pvLl}{L} 
\newcommand{\pvLv}{\vec{\pvLl}} 
\newcommand{\bvLl}{\ell} 
\newcommand{\bvLv}{\bivec{\bvLl}} 
\newcommand{\pvMl}{M} 
\newcommand{\pvMv}{\vec{\pvMl}} 
\newcommand{\bvMl}{m} 
\newcommand{\bvMv}{\bivec{\bvMl}} 
\newcommand{\pvNl}{N} 
\newcommand{\pvNv}{\vec{\pvNl}} 
\newcommand{\bvNl}{n} 
\newcommand{\bvNv}{\bivec{\bvNl}} 

\newcommand{\vUl}{U} 
\newcommand{\vUv}{\vec{\vUl}} 
\newcommand{\vVl}{V} 
\newcommand{\vVv}{\vec{\vVl}} 
\newcommand{\vWl}{W} 
\newcommand{\vWv}{\vec{\vWl}} 

\newcommand{\idxa}{a}
\newcommand{\idxb}{b}
\newcommand{\idxc}{c}

\newcommand{\xh}{\hat{x}}
\newcommand{\yh}{\hat{y}}
\newcommand{\zh}{\hat{z}}

\newcommand{\mfrac}[2]{\tfrac{#1}{#2}}
\newcommand{\mfracpd}[2]{\tfrac{#1}{#2}}
\newcommand{\mfracopts}[2]{#1}

\begin{document}

\title{Teaching Magnetism with Bivectors}

\author{Steuard \surname{Jensen}}
\email{jensens@alma.edu} 
\affiliation{Department of Physics and Engineering, Alma College, Alma, MI 48801}

\date{\today}

\begin{abstract}
The magnetic field is traditionally presented as a (pseudo)vector quantity, tied closely to the cross product. Though familiar to experts, many students find these ideas challenging and full of subtleties. Building on earlier work in rotational physics, we present an alternative pedagogical approach that describes magnetic fields using \textit{bivectors}. These objects can be visualized as oriented tiles whose components form an antisymmetric matrix. Historically, bivectors have been mostly used in specialized contexts like spacetime classification or geometric algebra, but they are not necessarily more complicated to understand than cross products. Teaching magnetism in this language addresses common student difficulties, generalizes directly to relativity (and extra dimensions), and brings fresh insight to familiar ideas.
\end{abstract}

\maketitle

\section{Introduction}
\label{sec:Introduction}

Learning magnetism for the first time is challenging. Students who have just started to understand the electric field are then introduced to a second field that is illustrated almost identically but behaves very differently. They must learn how and when to use various right-hand rules, which are known to be difficult for new learners. There are no two-dimensional examples which might serve as a simple starting point: every magnetism problem requires three-dimensional reasoning. Even after overcoming those challenges, more advanced students may still stumble over the magnetic field's pseudovector behavior or be puzzled by relativistic electromagnetism where electric and magnetic field components are unexpectedly jumbled into an antisymmetric tensor. 

\begin{figure}
\centering
\includegraphics[width=\linewidth]{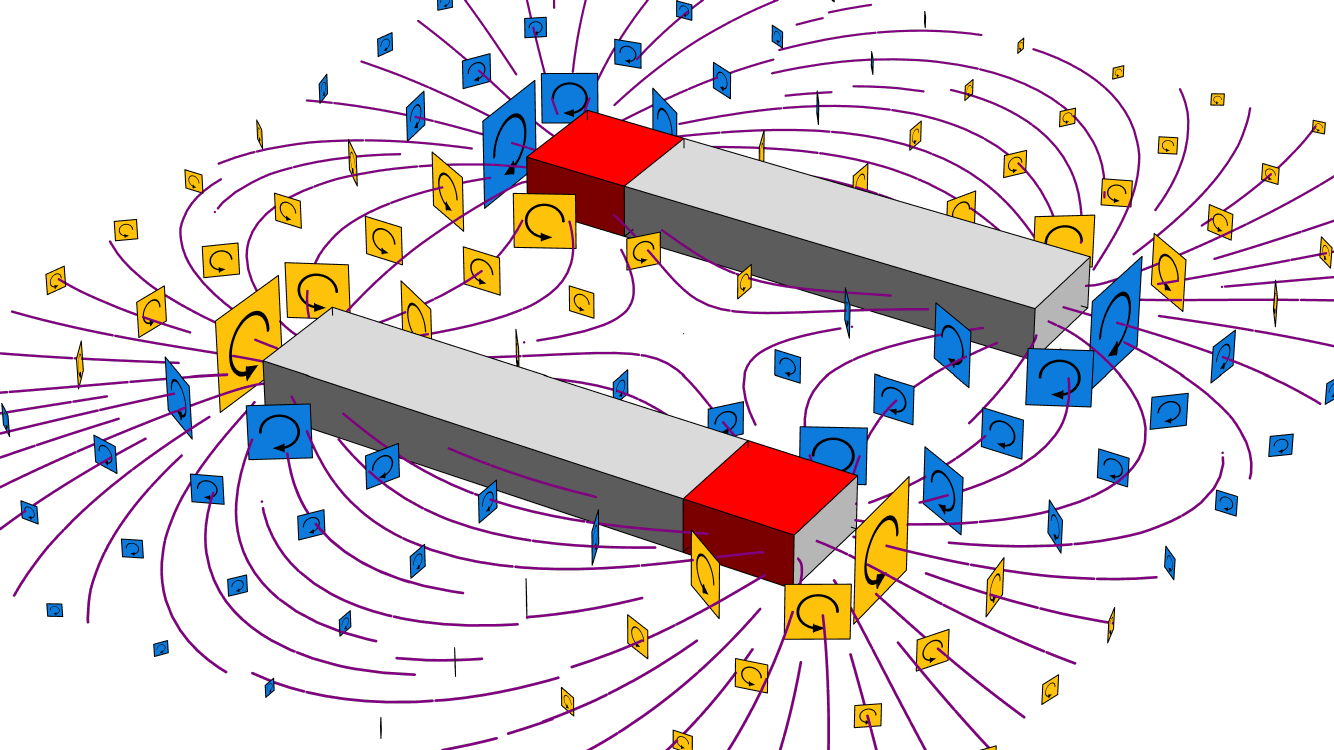} 
\caption{\label{fig:barMagnetBivecs}
Two bar magnets with north poles marked with tape. The resulting magnetic field lines are illustrated in the magnets' plane, along with bivector tiles representing the magnetic field. (The tiles might be imagined as ``beads'' threading along ``strings'' of field lines with each tile orthogonal to its line.) Tile orientation is labeled with an arrow: counterclockwise when looking into a north pole and clockwise when looking into a south pole. Larger tiles represent stronger fields. For visual clarity, the tiles are shaded so the counterclockwise side is lighter (orange) and the clockwise side is darker (blue).}
\end{figure}

Much of this becomes easier if we present the magnetic field as a \defn{bivector} quantity, instead of using the traditional vector description.\cite{Jancewicz:1980eb,Jancewicz:1988me} 
Figure~\ref{fig:barMagnetBivecs} illustrates two bar magnets and the resulting bivector magnetic field, visualized as oriented ``tiles'': a clear visual distinction from the vector electric field. 
This figure demonstrates that magnetic field lines can still be useful in the bivector formalism: we just label the field itself by tiles orthogonal to a line rather than by arrows along it.
Each tile's area represents the field's magnitude; the convention for the orientation (the arrow direction) is such that the field emerging from a north pole appears counterclockwise (when looking through the tile at the north pole).

As will be explained, the bivector formalism itself does not require the use of any right-hand rules%
\footnote{A simple right-hand rule can, however, relate each tile to the direction of the corresponding traditional vector.}
and avoids the subtleties of pseudovectors. Purely two dimensional examples are natural, because magnetic forces act in the bivector's plane, and the generalization to magnetism in four dimensional space-time becomes entirely straightforward.

Just as traditional instruction in magnetism often builds on prior experience using cross products to analyze rotational motion, teaching magnetism with bivectors follows naturally when students have previously studied rotations in that language. 
The author and a student discussed how to teach rotational physics using bivectors in previous work.\cite{Jensen:2022rot} (The present article can stand alone, but readers may prefer to first familiarize themselves with bivectors in that simpler context.) 
One encouraging message emerges in both contexts: instructors who prefer to emphasize the traditional vector descriptions may still benefit from incorporating some elements of bivectors.

\medskip

Even though bivectors are rarely discussed in the traditional curriculum, they have broad applicability: any equation involving cross products or curls can be rewritten in bivector language.
Most physicists have never studied bivectors, and it may at first seem unwise to present these unfamiliar ideas to students. But to new learners, cross products and their mathematics are just as unfamiliar and daunting: research shows significant struggles even after instruction.\footnote{Some of this is summarized in our previous paper,\cite{Jensen:2022rot} and Kustusch gives an excellent review of this literature.\cite{Kustusch:2016})}


Meanwhile, describing the magnetic field in the same vector language as the electric field has its own costs. 
Students very often analyze magnetic field problems using electric-like mental models (e.g.\ treating north/south poles like positive/negative charges).\cite{Guisasola:2004diff,Maloney:2001em,Scaife:2010dir} Furthermore, after studying magnetism, many students became confused about electric fields as well.\cite{Scaife:2011int}
A clear geometrical distinction between the descriptions of electric and magnetic fields can make these confusions less likely.

Mathematically, bivectors are rank-2 contravariant antisymmetric tensors.
(In fact, as discussed below the electromagnetic field tensor $F^{\mu \nu}$ is a (four-dimensional) bivector, though the term is rarely used.)
As such, bivectors are closely related to differential 2-forms (just as vectors are related to 1-forms). 
Rigorous studies of electromagnetism describe $E$ as a 1-form and $B$ as a 2-form field.\cite{vanDantzig:1934fe,Burke:1983mp}
Introductory treatments can be found in~\cite{Warnick:1997te,Fumeron:2020is}, with more detailed treatments appearing in (e.g.)~\cite{Hehl:2003fce,Burke:1985ag}.

Though they are elegant, differential forms are mathematically subtle and require a reformulation of not just magnetism but nearly everything. The diagrams used to visualize forms are also much less intuitive than vector arrows.\cite{Schouten:1954,MTW:1973}
By contrast, treating the magnetic field as a bivector is a single change to the traditional approach, 
and unlike differential forms the geometrical interpretation of bivectors relates straightforwardly to that of vectors.

There are a handful of other uses of bivectors in physics, including the Petrov classification of spacetimes\cite{Petrov:1954,Petrov:2000bs} and the study of general relativity with torsion.\cite{Trautman:2006fp}
Bivectors are also essential in geometric algebra, another sweeping mathematical reformulation of physics.\cite{Gull:1993,Hestenes:2003,Doran:2003,Vold:1993rot,Vold:1993EM} 
The present work advocates for the adoption of bivectors within an otherwise traditional formulation of physics, but learning these ideas may be helpful to those who later go on to study 
differential forms or geometric algebra.


The organization of this paper is as follows. Section~\ref{sec:LoopField} uses a current loop as context to introduce the bivector magnetic field. 
Section~\ref{sec:SourcesWedge} calculates magnetic fields from moving source charges, and
Section~\ref{sec:MagneticForces} then computes the forces exerted by magnetic fields on particles and currents.
Relativistic electromagnetism is discussed in Section~\ref{sec:Relativity}.
Finally, Section~\ref{sec:Conclusions} presents conclusions and notes on pedagogy.
Appendix~\ref{sec:MagneticTorque} discusses magnetic moment and torque.
Appendix~\ref{sec:GeomMatrixProd} presents geometric intuition for the matrix product. Appendix~\ref{sec:Advanced} sketches more advanced material including the vector potential, energy and momentum, extra dimensions, and field classification.
Last of all, Appendix~\ref{sec:BivectorReference} provides a reference for bivector operations.

\begin{figure}
\centering
\includegraphics[width=0.45\linewidth]{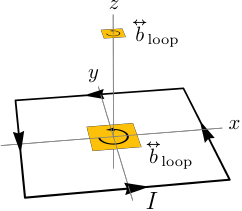}
\hfill
\includegraphics[width=0.45\linewidth]{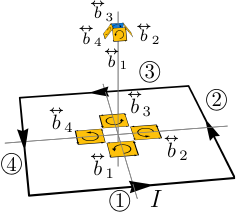}
\caption{\label{fig:SquareLoopMagFields} 
\textit{Left:} Inside a square current loop, the orientation of the produced bivector magnetic field matches the direction of current flow. Along the central axis the field weakens with distance, but its orientation remains the same.
\textit{Right:} The total field at any point is the sum of contributions from all four straight wires. In the plane of the loop, each side contributes a tile lying in that plane. 
Elsewhere, each contributed tile is tilted: one edge points toward the source, and the other is parallel to the current.
}
\end{figure}

\section{Essential intuition: the magnetic field of a current loop}
\label{sec:LoopField}


Regardless of formalism, instructors must choose to introduce magnetism with either the \emph{sources} of the magnetic field or with the field's \emph{effects}. Though either order can work, we choose to begin with an intuitive description of magnetic fields produced by flowing currents.

Figure~\ref{fig:SquareLoopMagFields} shows a geometric visualization of bivector fields. The bivector magnetic field $\bivec{b}$ at a point is represented as a two-dimensional tile whose \defn{area} is proportional to the bivector's magnitude and which has a specific \defn{attitude} (the angle of the plane in space) and \defn{orientation} (clockwise or counterclockwise).
The \emph{shape} of the tile is not physically significant, and in fact it is sometimes helpful to change from one shape to another.

The example of a square current loop in Fig.~\ref{fig:SquareLoopMagFields} displays an essential piece of intuition: the bivector magnetic field $\bivec{b}$ produced within an isolated (planar) current loop always has the same orientation as the flow of current around the loop. At points inside the loop, the contributions from each individual straight wire segment also have that same orientation.

This second observation is enough to determine the attitude and orientation of a straight wire's field contribution at \emph{any} point: just picture that wire as one edge of an imaginary (planar) loop encircling the measurement point. Then the field tile must be tilted so that it lies in the plane formed by the measurement point and the wire segment. For example, in Fig.~\ref{fig:SquareLoopMagFields}, the tile representing the contribution of wire~4 at a point above the loop is tilted down toward its source wire. 
The direction of the tile's orientation along the edge nearest the wire matches the direction of current flow.

\begin{figure}
\centering
\includegraphics[width=4cm]{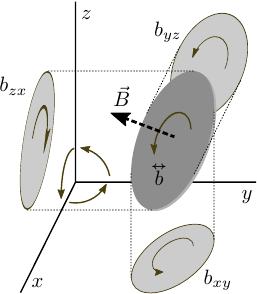}
\caption{\label{fig:tileProjections} The tile representing $\bivec{b}$ is projected 
onto each of the three coordinate planes. 
In this example we can see that $b_{xy}$ and $b_{yz}$ are positive while $b_{zx}$ is negative by comparing the orientation of each projection to the plane's orientation (conventionally chosen in cyclic order, and denoted by curved arrows near the origin).
If needed, to relate this to the equivalent traditional magnetic field, $\vec{B}$ is normal to the tile and derives its direction from the tile's orientation by the right-hand rule.}
\end{figure}

Like vectors, bivectors can be quantitatively described by their coordinate components, as illustrated in Fig.~\ref{fig:tileProjections}. Because tiles are two-dimensional those components correspond to \emph{planes} (each specified by two coordinate labels). For example, the magnitude of the component $b_{xy}$ is the area of the tile's projection on the $xy$ plane; the sign of that component is defined to be positive because the projection's orientation matches rotation of $x$ toward $y$. But since this is opposite the rotation of $y$ into $x$, we say that $b_{yx} = -b_{xy}$: the component labels are antisymmetric. 
Any component with a repeated label (like $b_{xx}$) is zero: it does not specify a plane.

Thus, specifying $b_{xy}$, $b_{yz}$, and $b_{zx}$ is enough to uniquely determine the bivector. (We have arbitrarily chosen to label these three coordinate pairs in cyclic order.)
Even though they label areas rather than lengths, these components combine according to the Pythagorean theorem to give the bivector's overall magnitude:\cite{Conant:1974gp}
\begin{equation}
\label{eq:bivecMagnitude}
|\bivec{b}| = b_{xy}^2 + b_{yz}^2 + b_{zx}^2
 = \frac{1}{2} \sum_{i,j = \{x,y,z\}} b_{ij}^2
 \quad.
\end{equation}
The factor of $\mfrac{1}{2}$ corrects for double counting, since (e.g.)\ $b_{xy}^2 = b_{yx}^2$. 
The components naturally form an antisymmetric matrix (or more precisely, a second rank tensor):
\begin{align}
\label{eq:bivecComponentIndices}
\bivec{b} &= 
\begin{pmatrix}
 0 & b_{xy} & b_{xz} \\
 b_{yx} & 0  & b_{yz} \\
 b_{zx} & b_{zy} & 0
\end{pmatrix}
= \begin{pmatrix}
 0 & b_{xy} & -b_{zx} \\
-b_{xy} & 0  & b_{yz} \\
b_{zx} & -b_{yz} & 0
\end{pmatrix}
\quad,
\end{align}
where in the second form we have put all component indices into cyclic order.

Figure~\ref{fig:tileProjections} also illustrates that the traditional magnetic field vector $\vec{B}$ is normal to the bivector $\bivec{b}$ in the direction given by a right-hand rule. 
The vector's components can be related to the bivector's as follows:
\begin{equation}
\label{eq:pseudovecBivecComponents}
\vec{B} = \begin{pmatrix} B_x \\ B_y \\ B_z \end{pmatrix}
 = \begin{pmatrix} b_{yz} \\ b_{zx} \\ b_{xy} \end{pmatrix}
 \quad.
\end{equation}
Because this equation relies on cyclic coordinate order, a right handed coordinate system becomes necessary any time we work with $\vec{B}$.



Finally, we have already discussed Fig.~\ref{fig:barMagnetBivecs}'s illustration of the magnetic field of two bar magnets. The bivector field orientation inside each magnet can be found using the rules already discussed in Sec.~\ref{sec:Introduction}, and together with our essential intuition about current loops this can connect naturally to the concept of bound currents.

\bigskip

\begin{figure}
\centering
\includegraphics[width=0.45\linewidth]{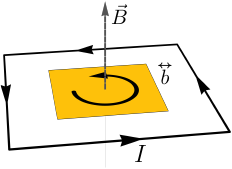}
\hfill
\includegraphics[width=0.45\linewidth]{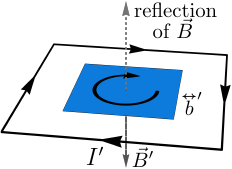}
\caption{\label{fig:LoopFieldReflection} 
\textit{Left:} A current loop's bivector magnetic field, along with its normal vector $\vec{B}$ (the traditional magnetic field).
\textit{Right:} The same scenario, reflected horizontally. The bivector orientation naturally reverses under reflection, but now the right-hand rule indicates that the corresponding $\vec{B}'$ should point down, while the reflection of the original $\vec{B}$ still points up. This extra reversal of $\vec{B}$ under reflection is the defining characteristic of a \defn{pseudovector}.}
\end{figure}

\begin{figure*}
\centering
\includegraphics[width=0.8\linewidth]{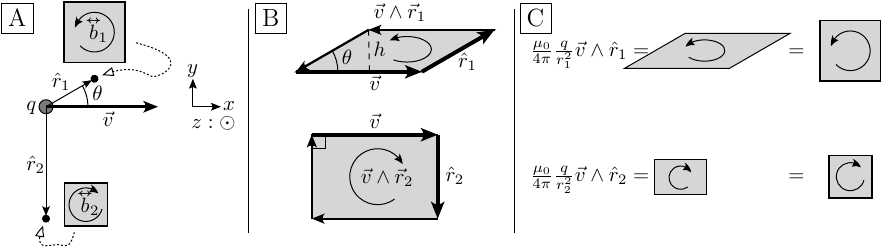}
\caption{\label{fig:ChargeWedgeTiles}%
\textit{A:} A positive point charge $q$ moving to the right with velocity $\vec{v}$ creates magnetic fields $\bivec{b}_{\!1}$ and $\bivec{b}_{\!2}$ (represented by tiles) at the two points shown. The tile orientation is counterclockwise for points above the line of motion and counterclockwise for points below it. 
\textit{B:} The calculation of each wedge product $\vec{v} \wedge \hat{r}$. Each vector pair defines a parallelogram in the plane of the page with area $|\vec{v} \wedge \hat{r}| = |\vec{v}| h = |\vec{v}| |\hat{r}| \sin\theta$. By placing the vectors tip to tail in product order (followed by their negatives to close the parallelogram), the arrows determine the tile's orientation. 
\textit{C:} Each field is computed from Eq.~\eqref{eq:PointChargeB}: a scalar multiplication simply rescales the area of the tile. For clarity, in the last step each tile has been reshaped to a square (while keeping its area and orientation the same).
}
\end{figure*}

We remind the reader that the traditional magnetic field vector $\vec{B}$ constructed by a right-hand rule is not a ``true'' vector (or \defn{polar vector}) at all, but rather a \defn{pseudovector} (or \defn{axial vector}). It behaves like a vector under rotations, but under a reflection its overall direction reverses (in addition to the reflection) as shown in Fig.~\ref{fig:LoopFieldReflection}. For example, under the reflection $x \to -x$, the components of $\vec{B}$ transform as $(B_x, B_y, B_z) \to (B_x, -B_y, -B_z)$. 
This awkward behavior becomes entirely natural in bivector components, where the sign changes only on the components involving $x$: $(b_{yz}, b_{zx}, b_{xy}) \to (b_{yz}, -b_{zx}, -b_{xy})$. This corresponds to reversing the signs of the $x$-row and $x$-column of the $\bivec{b}$ matrix (matching the usual tensor transformation law).

%
%

\section{Magnetic fields from moving charges and the wedge product}
\label{sec:SourcesWedge}

With the basic mathematics of bivectors established, we are ready for quantitative calculations.
Consider a point charge $q$ 
moving at constant (nonrelativistic) velocity $\vec{v}$. 
This scenario is illustrated in Fig.~\ref{fig:ChargeWedgeTiles}A, where the magnetic field bivector is computed at two points: one ahead of the charge at an angle $\theta$ from its direction of motion and one farther away directly to its side.
The equation for computing this bivector field is
\begin{equation}
\label{eq:PointChargeB}
\bivec{b} = \frac{\mu_0}{4 \pi} \frac{q}{r^2} (\vec{v} \wedge \hat{r}) \quad,
\end{equation}
where $\mu_0 \approx \qty[parse-numbers=false]{4\pi \times 10^{-7}}{\tesla \second \meter \per \coulomb}$, $\vec{r}$ is the position of the measurement point relative to the source, $r \equiv |\vec{r}|$, 
and $\hat{r} \equiv \mfrac{\vec{r}}{r}$ is a unit vector in that direction.

Instead of the traditional cross product we see the \defn{wedge product} of the vectors $\vec{v}$ and $\hat{r}$. As illustrated in Fig.~\ref{fig:ChargeWedgeTiles}B, the wedge product of two vectors defines a bivector. The tile representing this bivector is found by aligning the two vector arrows tip to tail in product order (followed by the same two vectors in reverse); the arrow directions define the bivector's orientation. This matches our essential intuition about bivector magnetic fields: the tile lies in the plane formed by $\vec{v}$ and $\hat{r}$, with the orientation of the near edge matching the flow of (positive) charge.

\begin{figure}
\centering
\includegraphics[width=0.8\linewidth]{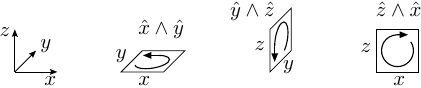}
\caption{\label{fig:CoordBivecs} 
Representations of unit bivectors in the given 3D coordinate system: e.g.\ $\hat{x}\wedge\hat{y}$ is a unit bivector in the $xy$-plane oriented in the direction that rotates $\hat{x}$ toward $\hat{y}$.}
\end{figure}

The magnitude of the wedge product bivector is the area of the parallelogram,
\begin{equation}
|\vec{v} \wedge \hat{r}| 
  \equiv |\vec{v}| \, |\hat{r}| \sin \theta = |\vec{v}| \sin \theta \quad.
\end{equation}
In the cases shown in Fig.~\ref{fig:ChargeWedgeTiles}A where the bivector tile lies in the plane of the page, we can write $\vec{v} \wedge \hat{r}_1 = |\vec{v}| \sin \theta \, \ccwbivec$ and $\vec{v} \wedge \hat{r}_2 = |\vec{v}| \sin \qty{90}{\degree} \, \cwbivec = |\vec{v}| \, \cwbivec$.
(The symbols $\ccwbivec$ and $\cwbivec$ indicate the orientation in the plane of the page 
and can be interpreted as ``unit bivectors''. Unit bivectors in other planes will be illustrated in Fig.~\ref{fig:CoordBivecs}.)

The wedge product is linear: $(a\vec{u}) \wedge \vec{v} = a (\vec{u} \wedge \vec{v}) = \vec{u} \wedge (a\vec{v})$ and $\vec{u} \wedge (\vec{v}+\vec{w}) = \vec{u} \wedge \vec{v} + \vec{u} \wedge \vec{w}$. It is also antisymmetric, $\vec{u} \wedge \vec{v} = -\vec{v} \wedge \vec{u}$: placing the two vectors tip to tail in the opposite order reverses the tile's orientation.
Consequently, the wedge product of a vector with itself has zero area: $\vec{v} \wedge \vec{v} = 0$. 

These results are also evident from the wedge product components:
\begin{equation}
\label{eq:WedgeComponents}
\bivec{\ell} = \vec{u} \wedge \vec{v}
\qquad \implies \qquad
\ell_{ij} = u_i v_j - u_j v_i
\quad,
\end{equation}
where $i,j=\{x,y,z\}$.
Expressing this in matrix form,
\begin{align}
\label{eq:WedgeMatrix}
\bivec{\ell} &= \begin{pmatrix}
0 & u_x v_y - u_y v_x & u_x v_z - u_z v_x \\
u_y v_x - u_x v_y & 0 & u_y v_z - u_z v_y \\
u_z v_x - u_x v_z & u_z v_y - u_y v_z & 0
\end{pmatrix} \\
\nonumber
 &= \begin{pmatrix}
0 & L_z & -L_y \\
-L_z & 0 & L_x \\
L_y & -L_x & 0
\end{pmatrix}
\quad.
\end{align}
The matrix components in the first line of Eq.~\eqref{eq:WedgeMatrix} are easy to remember: the subscripts of each positive term simply label the row and column (in that order) with no need to puzzle through cyclic coordinate patterns.
In the second line, we used Eq.~\eqref{eq:pseudovecBivecComponents} to replace the components of $\bivec{\ell}$ with the components of its corresponding normal (pseudo)vector $\vec{L}$. 
These component equations precisely match those for the cross product%
\footnote{In fact, this can be taken as the \emph{definition} of the cross product.}
$\vec{L} = \vec{u}\times\vec{v}$.

Rather than listing components, we can also use the wedge product to express quantities in terms of the coordinate unit vectors as illustrated in Fig.~\ref{fig:CoordBivecs}:  ``$\hat{x}\wedge\hat{y}$'' means ``a unit tile in the $xy$-plane, oriented in the direction that rotates $\hat{x}$ toward $\hat{y}$.'' 
In Fig.~\ref{fig:ChargeWedgeTiles}B, $\vec{v} \wedge \hat{r}_1 = |\vec{v}| \sin \theta \, \hat{x}\wedge\hat{y}$ and $\vec{v} \wedge \hat{r}_2 = |\vec{v}| \, \hat{x}\wedge(-\hat{y}) = -|\vec{v}| \, \hat{x}\wedge\hat{y} = |\vec{v}| \, \hat{y}\wedge\hat{x}$.

\medskip

Returning to Eq.~\eqref{eq:PointChargeB}, 
the magnetic field bivector is a scalar multiple of $\vec{v} \wedge \hat{r}$. This simply scales the magnitude of the bivector, and with it the area of the tile. In the final step shown in Fig.~\ref{fig:ChargeWedgeTiles}C, we have reshaped both tiles to squares for easier comparison of magnitudes (while keeping their areas and orientations the same). 
In explicit components, $b_{1,xy} = 
\mfracopts{\frac{\mu_0 q}{4\pi r_1^2} \, |\vec{v}| \sin\theta}%
{\mu_0 q |\vec{v}| \sin\theta/(4\pi r_1^2)}$ 
and $b_{2,xy} = 
-\mfracopts{\frac{\mu_0 q}{4\pi r_2^2}\,|\vec{v}|}%
{\mu_0 q |\vec{v}|/(4\pi r_2^2)}$.



\medskip

Now that we can compute fields due to point sources, we can address magnetostatics problems. For a steady current through a thin continuous wire composed of infinitesimal segments $d\vec{l}$, the magnetic field is computed as a sum using the Biot-Savart law:
\begin{equation}
\label{eq:BiotSavart}
\bivec{b} = \frac{\mu_0 I}{4 \pi} \int \frac{1}{r^2} (d\vec{l} \wedge \hat{r})
\quad.
\end{equation}
Sums and integrals of bivectors can be computed component by component just as for vectors.\footnote{Understanding bivector sums geometrically without resorting to components can be subtle: see~\cite{Jensen:2022rot}, Sec.~IV.}

\begin{figure}
\centering
\includegraphics[width=0.6\linewidth]{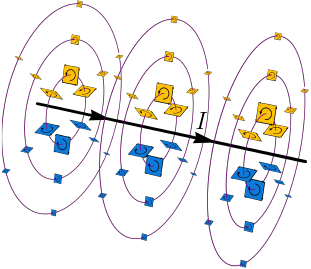}%
\caption{\label{fig:LineMagFieldBivec} 
An infinite straight wire carrying current $I$ produces the bivector magnetic field shown: in keeping with the Biot-Savart law's $I d\vec{l} \wedge \hat{r}$ term, each tile has an edge parallel to the wire source and another that extends radially away from it. The magnetic field lines circling around the wire are also shown.}
\end{figure}

For example, as illustrated in Fig.~\ref{fig:LineMagFieldBivec}, consider a current $I$ flowing through an infinite wire along the $x$-axis: $\vec{I} = I \hat{x}$. We seek to calculate the magnetic field at position $\vec{\rho} = y \hat{y} + z \hat{z}$, a distance $\rho = \sqrt{y^2 + z^2}$ from the wire. Then $d\vec{l} = dx \hat{x}$ and $\hat{r} = (x\, \hat{x} + y \,\hat{y} + z \,\hat{z})/r$, so
\begin{align}
\nonumber
\bivec{b} &= \frac{\mu_0 I}{4 \pi}
  \left( y \,\hat{x} \wedge \hat{y} + z  \,\hat{x} \wedge \hat{z} \right)
  \int _{-\infty}^\infty dx\, \frac{1}{(x^2+\rho^2)^{3/2}} \\
\label{eq:InfLine}
  &= \frac{\mu_0 I}{4 \pi}
  \left( \hat{x} \wedge \vec{\rho} \right) \frac{2}{\rho^2}
  = \frac{\mu_0 I}{2 \pi \rho} \, \hat{x} \wedge \hat{\rho}
 \;.
\end{align}
This agrees with the familiar vector result for the magnitude, and as usual each tile lies in a plane that extends both along the wire's direction and radially away from it. (To apply our essential intuition from Sec.~\ref{sec:LoopField}, we might imagine the long straight wire as part of a very large loop around any given measurement point.)
Naturally, we could instead have taken separate integrals for $b_{xy}$ and $b_{xz}$.

If the straight wire has a finite length $2L$, a similar calculation gives the field along the wire's central plane:
\begin{align}
\bivec{b} 
\label{eq:FiniteLine}
  &= \frac{\mu_0 I}{2 \pi \rho}  \frac{L}{\sqrt{\rho^2 + L^2}} \, \hat{x} \wedge \hat{\rho}
\quad.
\end{align}
  

\medskip

At last, we can return to our initial example illustrated in Fig.~\ref{fig:SquareLoopMagFields}: a loop of wire in the $xy$-plane forming a square of side length $2L$ and carrying steady current $I$. The field along the loop's central symmetry axis (the $z$-axis) is found by summing the contributions of the four segments. The field from wire~1 (for which $\vec{\rho} = L \hat{y} + z \hat{z}$) is:
\begin{align}
\bivec{b}_1  &= C \, \hat{x} \wedge (L \hat{y} + z \hat{z})
 \quad,
\end{align}
where the prefactor is
\begin{equation}
C  = \frac{\mu_0 I}{2 \pi (z^2 + L^2)}  \frac{L}{\sqrt{z^2 + 2L^2}} 
 \quad.
\end{equation}

The contributions of the other three wires are calculated similarly: 
\begin{equation}
\begin{aligned}
\bivec{b}_1 &= C\,  \hat{x} \wedge (\phantom{-{}}L \hat{y} + z \hat{z}) &
\bivec{b}_3 &= C\,  (-\hat{x}) \wedge (-L \hat{y} + z \hat{z}) \\
\bivec{b}_2 &= C\,  \hat{y} \wedge (-L \hat{x} + z \hat{z}) &
\bivec{b}_4 &= C\,  (-\hat{y}) \wedge (\phantom{-{}}L \hat{x} + z \hat{z})
\end{aligned}
\end{equation}
Rearranging the wedge products to a common order and summing the fields, the $\hat{x} \wedge \hat{y}$ components all reinforce each other, and the others cancel out. Overall,
\begin{equation}
\bivec{b}_\text{loop} = \frac{\mu_0 I}{2 \pi (z^2 + L^2)}  \frac{4L^2}{\sqrt{z^2 + 2L^2}} \,
  \hat{x} \wedge \hat{y}
\quad.
\end{equation}
This establishes our central result and intuition.

\section{Magnetic forces and the matrix product}
\label{sec:MagneticForces}

\begin{figure}
\centering
\includegraphics[width=0.32\linewidth]{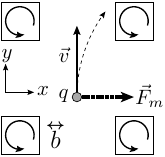}%
\caption{\label{fig:ChargeDeflectPlane} 
A positively charged particle moving with velocity $\vec{v}$ in a uniform magnetic field $\bivec{b}$ oriented counterclockwise in the plane. The direction of the magnetic force is found by rotating the velocity vector by \qty{90}{\degree} \emph{opposite} the field's orientation (here, a clockwise rotation). The particle will follow a clockwise circular path: this is opposite to the field's orientation as well.}
\end{figure}

Now that we understand the sources of magnetic fields in bivector language, we can study their effect on moving charges. Consider the example illustrated in Fig.~\ref{fig:ChargeDeflectPlane}: a particle of charge $q=+\qty{1}{\micro\coulomb}$ moving at $\qty{1}{\meter\per\second}$ in the $+y$ direction through a region of uniform magnetic field $\bivec{b} = \qty{6}{\milli\tesla}\,\ccwbivec$. 
(Equivalently, we could express the field in terms of its non-vanishing components $b_{xy} = -b_{yx} = \qty{6}{\milli\tesla}$.)

The bivector equation for the magnetic force on this particle involves an (order dependent!) ``dot product'' of the magnetic field and the velocity:
\begin{align}
\label{eq:magForce}
\vec{F}_m 
 &= q\, \bivec{b} \cdot \vec{v} \\
\label{eq:magForceMag}
 |\vec{F}_m| &= |q| \, |\bivec{b}| \, |\vec{v}| \cos\theta 
 \quad.
\end{align}
The second line gives the force's magnitude for general directions of motion, where $\theta$ is the angle between the velocity and the bivector plane.%
\footnote{Eq.~\eqref{eq:magForceMag} matches the familiar cross product formula: the traditional field vector $\vec{B}$ is normal to the bivector tile, so the angle between $\vec{B}$ and $\vec{v}$ is $\qty{90}{\degree}-\theta$, and $\cos\theta = \sin(\qty{90}{\degree}-\theta)$.}

The dot product in Eq.~\eqref{eq:magForce} formally denotes tensor index contraction, most conveniently interpreted as a matrix product with $\vec{v}$ as a column vector. 
Unlike the dot product of two vectors, this product is antisymmetric: $\bivec{b}\cdot\vec{v} = - \vec{v}\cdot\bivec{b}$. The geometric intuition behind the antisymmetry is illustrated in Appendix~\ref{sec:GeomMatrixProd},%
\footnote{\label{footnote:BivecDotIdentity}%
In brief, express $\bivec{b} = \vec{u} \wedge \vec{w}$ and compare two identities. 
With $\vec{v}$ on the left, $\vec{v} \cdot (\vec{u}\wedge\vec{w})
   = (\vec{v}\cdot\vec{u}) \,  \vec{w} - (\vec{v} \cdot \vec{w}) \, \vec{u}$:
the component of $\vec{v}$ along $\vec{u}$ gets rotated to point along $\vec{w}$ and the component along $\vec{w}$ gets rotated toward $-\vec{u}$: both components rotate in the same sense as the bivector's orientation. 
But $(\vec{u}\wedge\vec{w}) \cdot \vec{v} =
 \vec{u} \, (\vec{w} \cdot \vec{v}) - \vec{w} \, (\vec{u}\cdot\vec{v})$ 
 does the opposite: both terms rotate opposite to the bivector's orientation.
 This is also the origin of the ``\qty{90}{\degree} opposite the orientation'' rule.%
}
or it can be verified in components using matrix multiplication (with $\vec{v}$ in $\vec{v}\cdot\bivec{b}$ expressed as a row vector).

Returning to our example, the velocity is in the plane of the bivector ($\theta=0$), making the magnitude
$|\vec{F}_m| = |q| \, |\bivec{b}| \, |\vec{v}|
 = \qty{6}{\nano\newton}$. To find force's direction, start from the velocity vector and rotate \qty{90}{\degree} \emph{opposite} the orientation of the bivector field. In our case, the force acts in the $+x$ direction, and the particle eventually follows a clockwise circular path (which is likewise opposite the field's orientation).

The same result can be expressed using matrix multiplication:
\begin{align}
\label{eq:magForcePlane}
\vec{F}_m 
 &= \qty{1}{\micro\coulomb}
    \begin{pmatrix} 0 & \qty{6}{\milli\tesla} \\ -\qty{6}{\milli\tesla} & 0 \end{pmatrix}
    \begin{pmatrix} 0 \\ \qty{1}{\meter\per\second} \end{pmatrix}
  = \begin{pmatrix} \qty{6}{\nano\newton} \\ 0  \end{pmatrix}
\;.
\end{align}
One benefit of bivector language is already evident: this is a purely two dimensional example! (The traditional formulation would require three dimensional reasoning since $B_z \ne 0$.)

If students are not comfortable with matrix multiplication, one can simply list formulas for each component (expanded from index notation, $F_{m,i} = q \sum_{j=\{x,y,z\}} b_{ij} v_j$). Including all three dimensions for reference,
\begin{equation}
\label{eq:dotProdComponents}
\begin{aligned}
F_{m,x} &= q ( b_{xx} v_x + b_{xy} v_y + b_{xz} v_z )
 = q (b_{xy} v_y - b_{zx} v_z) \, , \\
F_{m,y} &= q ( b_{yx} v_x + b_{yy} v_y + b_{yz} v_z )
 = q (b_{yz} v_z - b_{xy} v_x) \, , \\
F_{m,z} &= q ( b_{zx} v_x + b_{zy} v_y + b_{zz} v_z )
 = q (b_{zx} v_x - b_{yz} v_y) \, .
\end{aligned}\end{equation}
Here, the antisymmetry of $\bivec{b}$ implies $b_{xx}=b_{yy}=b_{zz}=0$, and (purely for ease of comparison to the usual cross product) it allows us to standardize the other components in cyclic order.
The pattern of matching index labels in each component equation is straightforward to explain, and may be easier to remember than the cyclic pattern for cross products.

We can also work out our example in the language of coordinate unit bivectors, where the velocity is $\vec{v} = \qty{1}{\meter\per\second} \, \hat{y}$ and the magnetic field is $\bivec{b} = \qty{6}{\milli\tesla} \, \hat{x}\wedge\hat{y}$. (We do not need a separate $-\qty{6}{\milli\tesla} \, \hat{y}\wedge\hat{x}$ term: by antisymmetry, it would be redundant.)
Then from Eq.~\eqref{eq:magForce},
\begin{align}
\nonumber
\vec{F}_m &
= \qty{1}{\micro\coulomb} \;\; (\qty{6}{\milli\tesla} \, \hat{x}\wedge\hat{y}) \cdot \qty{1}{\meter\per\second} \,\hat{y} \\
\label{eq:forceEgbv}
 &= \qty{6}{\nano\newton} \, (\hat{x}\wedge\hat{y}) \cdot \hat{y}
   = \qty{6}{\nano\newton} \, \hat{x}
\quad.
\end{align}
In the last step we have used an identity in footnote~\ref{footnote:BivecDotIdentity}
to simplify the products of unit vectors. (A handy mnemonic for coordinate vectors only is that $\hat{y} \cdot \hat{y} = 1$ and $\hat{x} \wedge 1 = \hat{x}$.)

The general strategy for a dot product of a coordinate unit vector with a coordinate unit bivector is as follows. First (if necessary) use the antisymmetry of the wedge product to put matching unit vectors next to each other. Then let the dot product eliminate these two and remove the wedge product. For example,
\begin{align}
\label{eq:VecBivecDotExample}
\hat{y} \cdot (\hat{x}\wedge\hat{y})
 &= \hat{y} \cdot (-\hat{y}\wedge\hat{x}) 
 =  \hat{y} \cdot \hat{y}\wedge(-\hat{x})
 = -\hat{x}
\quad.
\end{align}
These examples illustrate the geometric meaning of the matrix product order: in Eq.~\eqref{eq:forceEgbv} (with the vector coming after the bivector) $\hat{y}$ is rotated \qty{90}{\degree} \emph{opposite} the bivector orientation, while in Eq.~\eqref{eq:VecBivecDotExample} (with the vector before the bivector) $\hat{y}$ rotates \qty{90}{\degree} \emph{with} the bivector orientation.

\begin{figure}
\centering
\includegraphics[width=0.67\linewidth]{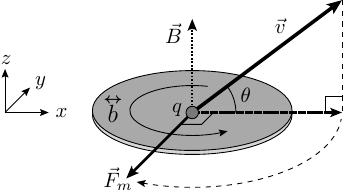}%
\caption{\label{fig:ChargeForce3D} 
A positively charged particle moving with velocity $\vec{v}$ at an angle $\theta$ from $\hat{x}$ toward $\hat{z}$ feels a force due to a magnetic field $\bivec{b}$ in the $xy$-plane. The direction of the magnetic force is found by first projecting the velocity vector into the plane of $\bivec{b}$ (reducing its length by a factor of $\cos\theta$) and then rotating it by \qty{90}{\degree} \emph{opposite} the field's orientation. The resulting magnitude is then $|\vec{F}_m| = |q|\,|\bivec{b}|\,|\vec{v}|\,\cos\theta$. 
The equivalent vector magnetic field $\vec{B}$ is shown for reference.
}
\end{figure}

\medskip

In general, the particle's velocity may lie outside the magnetic field's bivector plane. Consider the case illustrated in Fig.~\ref{fig:ChargeForce3D}, where our particle of charge $q=\qty{1}{\micro\coulomb}$ is moving through a magnetic field $\bivec{b} = \qty{6}{\milli\tesla}\, \hat{x}\wedge\hat{y}$ with velocity $\vec{v} = (4 \hat{x} + 3 \hat{z})\, \unit{\meter\per\second}$ (at an angle of $\theta \approx \qty{37}{\degree}$ above the $xy$-plane). Using Eq.~\eqref{eq:magForceMag}, we can compute the magnitude of the resulting magnetic force:
\begin{align}
|\vec{F}_m| &= \qty{1e-6}{\coulomb} \cdot
  \qty{6e-3}{\tesla} \cdot \qty{5}{\meter\per\second} \cdot \tfrac{4}{5}
 = \qty{24}{\nano\newton}
\quad.
\end{align}
The direction is found by first projecting the velocity vector into the bivector's plane 
and then rotating it \qty{90}{\degree} opposite the bivector's orientation. In this case, the resulting force points in the $-\hat{y}$ direction. 
 
Once again this can be expressed in matrix language:
 \begin{align}
\label{eq:magForce3D}
\vec{F}_m 
 &= \qty{1}{\micro\coulomb}
    \begin{pmatrix} 0 & \qty{6}{\milli\tesla} & 0 \\
     -\qty{6}{\milli\tesla} & 0 & 0 \\
     0 & 0 & 0 \end{pmatrix}
    \begin{pmatrix} \qty{4}{\meter\per\second} \\ 0 \\ \qty{3}{\meter\per\second} \end{pmatrix}
  = \begin{pmatrix} 0 \\ -\qty{24}{\nano\newton} \\ 0  \end{pmatrix}
.
\end{align}
 The explicit component expressions of Eq.~\eqref{eq:dotProdComponents} lead to the same result. In terms of coordinate unit vectors,
\begin{align}
\label{eq:magForce3Dhats}
\vec{F}_m &
= \qty{1}{\micro\coulomb} \;\; \qty{6}{\milli\tesla} \, \hat{x}\wedge\hat{y} \cdot (4 \hat{x} + 3 \hat{z})\, \unit{\meter\per\second} \\
\nonumber
 &= \qty{24}{\nano\newton} \, (\hat{x}\wedge\hat{y} \cdot \hat{x})
         + \qty{18}{\nano\newton} \, (\hat{x}\wedge\hat{y} \cdot \hat{z})
   = -\qty{24}{\nano\newton} \, \hat{y}
\;.
\end{align}
In the second term, $\hat{z}$ is orthogonal to both $\hat{x}$ and $\hat{y}$, and thus the dot product is zero.
 
\bigskip

Other standard results for magnetic forces translate naturally into bivector language. For example, the conclusion that magnetic forces do no work follows directly from the antisymmetry of $\bivec{b}$:
\begin{equation}
W = \int \vec{F} \cdot d\vec{l}
  = \int (-q \vec{v} \cdot \bivec{b}) \cdot (\vec{v} \,dt)
  = -q \int (\vec{v} \cdot \bivec{b} \cdot \vec{v} ) \,dt
  = 0 
\;.
\end{equation}
Similarly, the magnetic force on a thin wire carrying current $I$ along a path formed from infinitesimal segments $d\vec{l}$ is given by the vector integral
 \begin{equation}
 \label{eq:MagForceWire}
 \vec{F}_m = \int I\, \bivec{b} \cdot d\vec{l} \quad.
 \end{equation}
For a straight wire of directed length $\vec{L}$ in a uniform field, this leads to the result $\vec{F}_m = I\, \bivec{b} \cdot \vec{L}$.

\section{Electromagnetism and relativity}
\label{sec:Relativity}

The bivector description of magnetism also gives fresh insight in four dimensional spacetime.%
\footnote{It is unusual to discuss relativity at the introductory level, but there are exceptions.\cite{Moore:SI4E}}
In special relativity, we extend our usual three-dimensional vectors by incorporating an additional time component to create a four-vector. Position $(x,y,z)$ combines with time $t$ as the spacetime position $\fourvec{X} = (ct, x, y, z)$; momentum $\vec{p}$ becomes four-momentum by adding energy as its time component: $\fourvec{P} = (E/c, p_x, p_y, p_z) = m \fourvec{U}$. This last form uses the relativistic four-velocity $\fourvec{U} = \mfrac{d\fourvec{X}}{d\tau} = (c\, \mfrac{dt}{d\tau},\mfrac{dx}{d\tau},\mfrac{dy}{d\tau},\mfrac{dz}{d\tau}) = \gamma\, (c,v_x,v_y,v_z)$, where $\tau$ is proper time and $\gamma = \mfrac{dt}{d\tau} = 1/\sqrt{1-v^2/c^2}$.

Electric and magnetic fields, however, do not extend to four-vectors. Instead, careful arguments show that relativistic transformation laws mix their components together. Imagine, though, that we had treated magnetic fields as bivectors from the start. Then the natural generalization of the $3\times3$ matrix for $\bivec{b}$ in~\eqref{eq:bivecComponentIndices} would be a $4\times4$ matrix, adding a new $t$-row and $t$-column at the top and left (superscripts denote contravariant indices):
\begin{align}
\fourvec{b}^{\mu \nu} &= 
\begin{pmatrix}
 0 & b^{tx} &b^{ty} & b^{tz} \\
 -b^{tx} & 0 & b^{xy} & -b^{zx} \\
 -b^{ty} & -b^{xy} & 0  & b^{yz} \\
 -b^{tz} & b^{zx} & -b^{yz} & 0
\end{pmatrix}
\quad.
\end{align}
Ordinary rotations do not mix the new time-space components with the familiar space-space components, but Lorentz boosts do.%
\footnote{Lorentz transformations (rotations and boosts) of four-vectors can be written as a matrix product: $\fourvec{U}' = \Lambda \fourvec{U}$. In that language, a four-bivector transforms as $\protect\bivec{\fourvec{b}}' = \Lambda \protect\bivec{\fourvec{b}} \Lambda^T$. These transformations precisely reproduce the Lorentz transformations of $\vec{E}$ and $\vec{B}$.} 
This corresponds precisely to the electromagnetic field tensor usually found in more advanced treatments:
\begin{align}
\fourvec{F}^{\mu \nu} &= 
\begin{pmatrix}
 0 & E_x/c & E_y/c & E_z/c \\
 -E_x/c & 0 & B_z & -B_y \\
 -E_y/c & -B_z & 0  & B_x \\
 -E_z/c & B_y & -B_x & 0
\end{pmatrix}
\quad.
\end{align}
The correspondence between the $\vec{B}$ and $\bivec{b}$ components is familiar, and now we see that the electric field emerges naturally from the four-bivector's timelike components.%
\footnote{This process of ``dimensional reduction'' is common in theories of extra dimensions: an antisymmetric tensor in $d$~dimensions splits into an antisymmetric tensor plus a vector when one direction is singled out and the system is described in $d-1$ dimensions.}

The rules for effects and sources of magnetic fields generalize directly to this relativistic form. We saw in Eq.~\eqref{eq:magForce} that the force on a moving particle was $\vec{F}_m = q \bivec{b} \cdot \vec{v}$, or using index notation, $F_{m,i} = q \sum_j b_{ij} v_j$. For relativistic four-vectors, this becomes
\begin{align}
\label{eq:FourForce}
\fourvec{F}_\text{em}^\mu &= \frac{d\fourvec{P}^\mu}{d\tau}
 = q \sum_\nu \fourvec{b}^{\mu \nu} \fourvec{U}_\nu
 = q \sum_{\nu, \rho} \fourvec{b}^{\mu \nu} \eta_{\nu \rho} \fourvec{U}^\rho 
\quad.
\end{align}
Here, $\eta_{\mu \nu} = \diag(-1,1,1,1)$ is the flat metric that defines the 
inner product. An example component is:
\begin{align}
F_\text{em}^x &= \frac{d\fourvec{P}^x}{dt} = \frac{d\fourvec{P}^x}{d\tau} \frac{d\tau}{dt}
 = \frac{d\fourvec{P}^x}{d\tau} \frac{1}{\gamma}
 = \frac{1}{\gamma} \fourvec{F}_\text{em}^x \\
\nonumber
 &= \frac{1}{\gamma} q \Bigl( \tfrac{E_x}{c} (\gamma c) + \sum_j b_{xj} (\gamma v_j) \Bigr)
 = q \bigl( E_x + \sum_j b_{xj} v_j \bigr)
,
\end{align}
which exactly matches the usual Lorentz force law. 

The components of the $4\times 4$ generalization of $\bivec{b}$ are sensible in the case of a non-relativistic point charge ($\gamma \approx 1$). Taking $\hat{\fourvec{R}} = (0,\hat{r})$, the natural generalization of Eq.~\eqref{eq:PointChargeB} is
\begin{equation}
\bivec{\fourvec{b}} = \frac{\mu_0}{4 \pi} \frac{q}{r^2} (\fourvec{U} \wedge \hat{\fourvec{R}}) \quad.
\end{equation}
The space-space components are the same as in the $3\times 3$ case, and the time-space components give $\fourvec{b}^{ti} = (\mfrac{\mu_0 c}{4\pi}) (\mfrac{q}{r^2}) \,\hat{r}^i$, or since $\epsilon_0 \mu_0 = 1/c^2$, $\vec{E} = \mfracpd{q}{4\pi\epsilon_0 r^2} \,\hat{r}$: as expected from Coulomb's law.

\medskip

\begin{figure}
\centering
\includegraphics[width=0.44\linewidth]{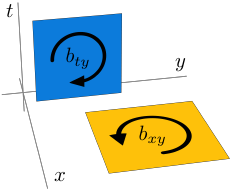}%
\hfill
\includegraphics[width=0.46\linewidth]{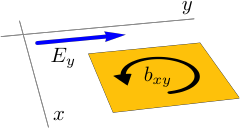}%
\caption{\label{fig:TilesTimeSpace} 
\textit{Left:} An EM field in spacetime is represented as a sum of a purely spacelike tile ($b_{xy}$) and a tile with one edge in the time direction ($b_{ty}$). \textit{Right:} Viewed as spatial fields, a magnetic field bivector in the $xy$-plane and an electric field vector in the $y$-direction are present.}
\end{figure}

It is enlightening to consider the geometric significance of $\bivec{\fourvec{F}}$. As discussed further in Appendix~\ref{sec:ExtraDims}, in $d$ dimensions it is possible to represent any bivector as a sum of $d/2$ or fewer tiles (rounded down). 
As illustrated in Fig.~\ref{fig:TilesTimeSpace}, in four-dimensional spacetime, it is natural to consider $\bivec{\fourvec{F}}$ as a sum of the usual spatial magnetic field tile (extended to a four-bivector with vanishing time components) plus an ``electric tile'' $\hat{\fourvec{t}} \wedge \fourvec{E}/c$, where $\hat{\fourvec{t}} = (1,0,0,0)$ and $\fourvec{E} = (0,E_x,E_y,E_z)$.
(The electric tile's orientation follows naturally from our essential intuition in Section~\ref{sec:LoopField} if we consider a stationary charge as a ``current'' in the time direction.)
In special cases, the field can be represented by a single (tilted) tile: identifying these cases is the subject of Appendix~\ref{sec:FieldClassification} which discusses the algebraic classification of field configurations.

\section{Pedagogy and Conclusions}
\label{sec:Conclusions}

The bivector description of magnetism is mathematically equivalent to the traditional (pseudo)vector approach; component by component, the equations are the same. We trade away the well-known vector operations (and the subtlety of the cross product) in favor of a less familiar picture that has more direct relationships to the field's sources and effects.

Teaching magnetism as a bivector helps students avoid the known pitfalls of the (pseudo)vector approach, but that has to be balanced against the costs of introducing the new formalism. Individual instructors can judge how much of the new picture to incorporate. 

A pure bivector treatment would entirely eliminate the need to teach right-hand rules, dwell on the cross product, or worry about left-handed coordinate systems. However, that is not practical today: students need to engage with the traditional approach in other courses and contexts. But, in the author's experience, students appreciate learning the direction of the cross product as ``first construct the wedge product tile, then curl your right-hand fingers in the direction of the tile's orientation: your thumb points in the direction of the cross product.'' It doesn't cost much class time to comment that the parallelogram tile itself may be a ``truer'' description of the magnetic field, and that we use the right-hand rule to convert it to a vector for computational convenience.

\begin{figure}
\centering
\includegraphics[width=0.8\linewidth]{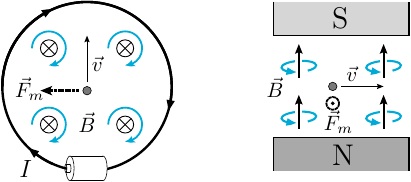}%
\caption{\label{fig:DecVectorExamples} 
Typical magnetism examples showing traditional vector fields $\vec{B}$  ``decorated'' with bivector orientation arrows, which are easier to draw than 3D bivector tiles. 
For the field inside a current loop, the orientation decorations provide an easy guide to the proper $\vec{B}$ direction.
And in each case, the decorations give students a visual cue that the field's effects are not along $\vec{B}$ but ``around'' it.}
\end{figure}

Students also tend to confuse electric and magnetic fields. Presenting magnetic fields primarily as tiles rather than arrows would make the distinction clear. But if that is too big a leap, instructors can draw magnetic fields as illustrated in Fig.~\ref{fig:DecVectorExamples}: the traditional (pseudo)vector with a curved arrow around it indicating the right-hand rule orientation of the corresponding bivector.\footnote{A similar convention is sometimes used by engineers to illustrate torque, another pseudovector quantity.}

These ``decorated'' vectors look at least somewhat distinct from electric field vectors, while retaining a direct visual relationship to the orientation of current loop sources. (Also, like bivector tiles, they may remind students that finding the force direction involves rotating in the curved arrow's plane.) With the right pedagogical framing, problems involving planar systems might be approachable using decorated vectors pointing into or out of the plane, without the need for much three-dimensional reasoning. 
This could have real benefits, possibly even at the high school level.

Of course, adopting a new formalism even in part is challenging on its own. The cross product may be confusing for students, but that one operation can be applied to any combination of vectors and pseudovectors being multiplied. (In principle, students using the cross product with decorated vectors would need to keep track of whether or not the result was a pseudovector.) With bivectors, it is necessary to distinguish between the product of two vectors in Eq.~\eqref{eq:WedgeComponents}, the product of a bivector and a vector in Eq.~\eqref{eq:magForce}, and the product of two bivectors in Eq.~\eqref{eq:MagTorqueComProd}. That is a lot for students to keep track of. In the author's experience, students respond positively to decorated vectors and bivector-inspired right-hand rules, but at the introductory level it is best to encourage them to use any right-hand rule that makes sense, without worrying whether it matches the proper bivector geometry.

Finally, the lack of direct support for bivectors in existing textbooks presents a challenge for instructors.  In the author's experience, providing a single clear story is better than explaining one approach first followed by the alternative, which comes across as repetitive and confusing.
Without textbooks or other sources that describe the magnetic field as a bivector, it is hard to present that as the primary description. 
Similarly, computational tools like Mathematica and Maple provide built-in support for vectors and vector fields, but handling bivectors requires more effort.%
\footnote{The author is happy to share the Mathematica code used to generate figures in this article, but it is not (yet) clean or robust enough for public consumption.}

Clearly, there is a tension between these two approaches. The bivector description may have moderate pedagogical advantages, while the pseudovector description has the benefit of over a century of development and tradition. It is natural to ask, if the two are mathematically equivalent, why go to the trouble to switch?

Magnetic field bivectors are true tensors, not awkward pseudovectors. 
Students who go on to study electromagnetism with differential forms (or geometric algebra) may benefit from this initial exposure to the magnetic field as a rank-2 tensor.
Relativistic electromagnetism has always required a bivector formulation, but traditionally students are given no geometric insight into that structure. Unlike with vectors, the bivector approach can describe a purely two-dimensional problem, as well as magnetic effects in theories of higher dimensions. 
For all these reasons, it is reasonable to conclude that bivectors are a more fundamentally \emph{true} description of magnetic fields, and we owe it to our students to share that with them.
Just as significant, the bivector approach brings fresh insight into this familiar topic by emphasizing the field to be both cause and effect of rotating charges.

We can at least open the door to these deeper ideas. Even if it is not feasible to fully embrace the bivector description right away, limited steps in this direction may have pedagogical benefits. There is more development to be done: most notably, a presentation of Maxwell's equations in bivector language is important, but requires a careful development of bivector calculus.\cite{Jensen:2024max} But these core ideas stand strongly on their own, and already address the substantial majority of topics in magnetism at the introductory level.

\begin{acknowledgments}
Thanks to Jack Poling for help developing these ideas in the context of rotational physics and for a careful literature search.
Figures~\ref{fig:barMagnetBivecs}, \ref{fig:SquareLoopMagFields}, \ref{fig:LoopFieldReflection}, \ref{fig:LineMagFieldBivec}, and \ref{fig:TilesTimeSpace} were produced with Mathematica.
Figures~\ref{fig:tileProjections} and \ref{fig:CoordBivecs} by the author and most of the reference information in Appendix~\ref{sec:BivectorReference} previously appeared in Ref.~\cite{Jensen:2022rot}.
\end{acknowledgments}



\appendix

\section{Current loops in external fields}
\label{sec:MagneticTorque}

A current loop in a uniform magnetic field $\bivec{b}$ experiences a torque, which is properly understood as a bivector quantity.\cite{Jensen:2022rot} To calculate this, we must first define the magnetic moment bivector of a current flowing along the closed curve $\mathcal{C}$: $\bivec{\mu} \equiv I \bivec{a}$. Here $\bivec{a}$ is the curve's directed area: for a planar loop, this is just a bivector whose magnitude is the area and whose orientation matches the flow of current around the curve.
More generally, $\bivec{a}$ is computed as a sum of infinitesimal bivector areas $d\bivec{a}$ that form a surface $\mathcal{S}$ bounded by closed curve $\mathcal{C}$:%
\footnote{The formula at the end of Eq.~\protect\eqref{eq:AreaBivec} results from choosing $\mathcal{S}$ as the cone with tip at the origin and $\mathcal{C}$ as its base. Each segment $d\vec{\ell}$ is the base of a triangle whose tip is at the origin and is exactly half of the bivector tile $\vec{r} \wedge d\vec{\ell}$.}
\begin{align}
\label{eq:AreaBivec}
\bivec{a} &= \int_\mathcal{S} d\bivec{a}
 = \frac{1}{2} \oint_\mathcal{C} \vec{r} \wedge d\vec{\ell} \quad, \; \text{or in components,} \\
\label{eq:AreaBivecComponents}
 a_{ij} &= \frac{1}{2} \oint_\mathcal{C} (r_i d\ell_j - r_j d\ell_i)
  = \oint_\mathcal{C} r_i d\ell_j
\quad.
\end{align}
Any surface $\mathcal{S}$ bounded by $\mathcal{C}$ will yield the same $\bivec{a}$. 
The final equality in Eq.~\eqref{eq:AreaBivecComponents} uses integration by parts, where $d\vec{\ell}$ is just $d\vec{r}$ restricted to lie along the curve.

We now consider the torque $d\bivec{\tau}$ acting on each segment $d\vec{\ell}$ of the wire. Infinitessimally, $d\bivec{\tau} = \vec{r} \wedge d\vec{F}$, where the force is given by Eq.~\eqref{eq:MagForceWire}. Then
\begin{align}
\bivec{\tau} &= \oint_\mathcal{C} \vec{r} \wedge (I\bivec{b} \cdot d\vec{\ell})
 = I \oint_\mathcal{C} \vec{r} \wedge (\bivec{b} \cdot d\vec{\ell})
\quad,
\end{align}
or after a few lines of straightforward index notation,%
\footnote{%
Each dot product is a matrix multiplication, or more precisely, tensor index contraction on one index. In this work, we will always denote this with a (single) dot product like $\protect\bivec{b} \cdot \protect\bivec{\mu}$. In geometric algebra the convention is different: there, roughly speaking, a (single) dot product denotes contraction on as many indices as possible. Here, when we need to contract both bivector indices as in Eq.~\protect\eqref{eq:DipoleEnergy}, our notation will be the ``double dot product''.} 
\begin{align}
 \label{eq:MagTorqueComProd}
 \bivec{\tau} &= \bivec{b} \cdot \bivec{\mu} - \bivec{\mu} \cdot \bivec{b}
  \equiv \bivec{b} \boxtimes \bivec{\mu}
\quad.
\end{align}
In this last step we have defined the notation $\boxtimes$ for the \defn{bivector cross product} or \defn{commutator product} of two bivectors producing a bivector.%
\footnote{The symbol $\boxtimes$ is meant to evoke a cross product applied to bivector tiles. The traditional notation for the commutator of two matrices is $\protect\bivec{\tau} = [\protect\bivec{b},\protect\bivec{\mu}]$, but a notation that looks like a product seems helpful.} 
(Note the reversed order compared with $\vec{\tau} = \vec{\mu} \times \vec{B}$.)

This bivector cross product shares many features with the familiar vector cross product, including antisymmetry: $\bvMv \boxtimes \bvLv = -\bvLv \boxtimes \bvMv$. Its geometric meaning is interesting (essentially a wedge product extended into a third dimension, creating a parallelogram tube by placing bivector tiles edge to edge: see Fig.~\ref{fig:TileCommute} in Appendix~\ref{sec:BivectorReference}) but sufficiently complicated that it's probably not worth sharing it with students.  For calculations, an identity%
\footnote{If $\protect\bvMv = \vec{A} \wedge \vec{B}$ and $\protect\bvLv = \vec{U} \wedge \vec{V}$, then $\protect\bvMv \boxtimes \protect\bvLv =
  (\vec{A} \wedge \vec{V}) (\vec{B} \cdot \vec{U})
    - (\vec{A} \wedge \vec{U}) (\vec{B} \cdot \vec{V}) 
    - (\vec{B} \wedge \vec{V}) (\vec{A} \cdot \vec{U})
    + (\vec{B} \wedge \vec{U}) (\vec{A} \cdot \vec{V})$.} 
allows us to simplify commutator products of unit bivectors:
\begin{equation}
\begin{aligned}
(\xh\wedge\yh) \boxtimes (\yh\wedge\zh) &= \xh \wedge \zh &
  (\xh\wedge\yh) \boxtimes (\zh\wedge\yh) &= -\xh \wedge \zh \\
(\xh\wedge\yh) \boxtimes (\zh\wedge\xh) &= \yh \wedge \zh &
  (\yh\wedge\zh) \boxtimes (\xh\wedge\zh) &= -\yh \wedge \xh \\
(\xh\wedge\yh) \boxtimes (\xh\wedge\yh) &= 0 &
 \text{etc.}
\end{aligned}
\end{equation}
The pattern is straightforward: reverse the wedge products as necessary to put the matching coordinate unit vectors next to each other, adding a minus sign for each reversal, and the result is the wedge product of the remaining two unit vectors.

Finally, the energy of a magnetic dipole $\bivec{\mu}$ (like a current loop) in a magnetic field $\bivec{b}$ can be expressed as
\begin{align}
\label{eq:DipoleEnergy}
U &= -\tfrac{1}{2} \bivec{\mu} \dblcdot \bivec{b}
 \equiv -\tfrac{1}{2} \sum_{i,j} \mu_{ij} b_{ij}
 = -|\bivec{\mu}| |\bivec{b}| \cos\theta  \\
\nonumber
 &= -(\mu_{xy} b_{xy} + \mu_{yz} b_{yz} + \mu_{zx} b_{zx})
\quad.
\end{align}
This is the ``double dot product'' of bivectors, often denoted by a colon, where $\theta$ is the angle between the two bivector planes.

\section{Geometric intuition for the matrix product}
\label{sec:GeomMatrixProd}

\begin{figure}
\centering
\vspace{1em}
\begin{overpic}[percent,width=0.667\linewidth]{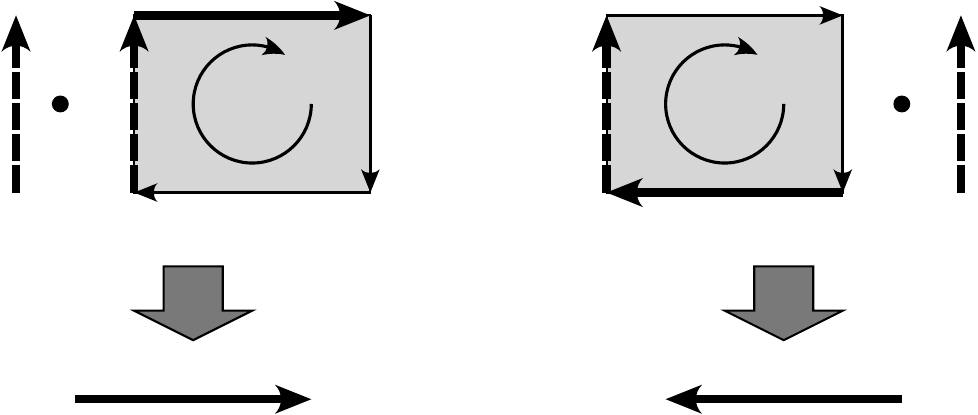}%
  \put(-4,29.5){$\vec{v}$}
  \put(23.5,29.5){$\bivec{b}$}
  \put(9,29.5){\color{gray}$\vec{u}$}
  \put(23.5,42.5){\color{gray}$\vec{w}$}
  \put(72,29.5){$\bivec{b}$}
  \put(57.5,29.5){\color{gray}$\vec{u}$}
  \put(69.25,17.75){\color{gray}$-\vec{w}$}
  \put(101,29.5){$\vec{v}$}
\end{overpic}
\caption{\label{fig:BivecDotVecOrders} 
An illustration of the antisymmetry \mbox{$\vec{v}\cdot\bivec{b} = -\bivec{b}\cdot\vec{v}$}. 
The dot product of a vector with a bivector ``absorbs'' the dimension along the vector. Here, the bivector tile's shape is chosen as a rectangle aligned with $\vec{v}$. The bold edges represent the terms in a wedge product: $\bivec{b} = \vec{u}\wedge\vec{w} = -\vec{w}\wedge\vec{u}$. When the dot product with $\vec{v}$ is on the left, we write $\bivec{b}$ so the vector parallel to $\vec{v}$ is adjacent to it: $\vec{v} \cdot (\vec{u}\wedge\vec{w})$. The parallel vectors are ``absorbed,'' leaving $\vec{w}$ to determine the result's direction. When the dot product with $\vec{v}$ is on the right, we instead write $(-\vec{w}\wedge\vec{u})\cdot \vec{v}$ and the result points along $-\vec{w}$.}
\end{figure}

Figure~\ref{fig:BivecDotVecOrders} illustrates the the geometric intuition behind the behavior and antisymmetry of the dot product of a bivector and a vector, $\vec{v}\cdot\bivec{b} = -\bivec{b}\cdot\vec{v}$. As described there, the dot product with $\vec{v}$ essentially ``absorbs'' away one dimension of the bivector (the one parallel to $\vec{v}$'s projection to the bivector plane), just as the dot product of a vector with a vector absorbs away its one direction to leave a scalar. A dot product on the left lets the bivector orientation carry it ``forward'' to yield a vector rotated \qty{90}{\degree} from $\vec{v}$. A dot product on the right carries $\vec{v}\,$'s direction ``backward'' by \qty{90}{\degree} instead, opposite the bivector's orientation.

If the bivector tile $\bivec{b} = \vec{u}\wedge\vec{w}$ is not so conveniently aligned with $\vec{v}$, a more general relationship follows from the component expression $b_{ij} = u_i w_j - u_j w_i$:
\begin{align}
\nonumber
(\vec{v} \cdot \vec{b})_j &=\sum_{i=1}^3 v_i \,(u_i w_j - u_j w_i) \\
 &= \sum_{i=1}^3 (v_i u_i)\, w_j - (v_i w_i)\, u_j \quad, \text{ so} \\
\vec{v} \cdot (\vec{u}\wedge\vec{w})
   &= (\vec{v}\cdot\vec{u}) \,  \vec{w} - (\vec{v} \cdot \vec{w}) \, \vec{u} \quad.
\end{align}
Any component of $\vec{v}$ oriented along $\vec{u}$ gets rotated to point along $\vec{w}$ and any component oriented along $\vec{w}$ gets rotated to point along $-\vec{u}$, matching the bivector orientation direction. A similar derivation gives
\begin{align}
(\vec{u}\wedge\vec{w}) \cdot \vec{v} &=
 \vec{u} \, (\vec{w} \cdot \vec{v}) - \vec{w} \, (\vec{u}\cdot\vec{v}) \quad,
\end{align}
which is exactly the negative of the previous result with exactly the opposite rotation. In the special case where $\vec{v} \cdot \vec{w} = 0$, these match the visualization in Figure~\ref{fig:BivecDotVecOrders}.

\section{More advanced material}
\label{sec:Advanced}

\subsection{The vector potential}
\label{sec:VecPotential}

Although the topic goes beyond the introductory material that has been the focus above, it is worth briefly noting how the magnetic vector potential $\vec{A}$ fits into the bivector formalism. The answer is straightforward: the vector potential remains an ordinary vector, and all of its usual properties and relationships remain unchanged.

The one necessary adjustment is its relationship to the magnetic field. Rather than the traditional curl $\vec{B} = \nabla \times \vec{A}$ we use the \defn{bivector curl} (or \defn{exterior derivative}):
\begin{align}
\bivec{b} &= \nabla \wedge \vec{A} \quad \qquad \text{or in components,} \\
b_{ij} &= \partial_i A_j - \partial_j A_i
\quad,
\end{align}
where $\partial_i$ is a shorthand for $\frac{\partial\phantom{A}}{\partial x_i}$.
Under the relationship defined in Eq.~\eqref{eq:pseudovecBivecComponents} these are precisely the components of the curl of $\vec{A}$, structured as a bivector.

The proper intuition for this bivector curl is precisely the same as for the usual curl, except that the resulting bivector orientation directly reflects the rotational quality of the vector field (rather than requiring a right-hand rule step to transform it to an axis of rotation). 
In that regard, the standard vector potential relationship $\vec{A} = \tfrac{\mu_0 I}{4\pi} \int (\tfrac{d\vec{\ell}}{r})$ (for currents that vanish at infinity) may give insight into the form of the bivector field due to moving charges. Roughly speaking, a moving charge drags $\vec{A}$ along with it, diminishing with distance. Thus, the charge produces a bivector curl $\bivec{b}$ in the current-distance plane, with its orientation paralleling the current on the nearer side.

\subsection{Energy and momentum of the field}
\label{sec:EnergyMomentum}

The expression for the energy density of the magnetic field is essentially the same in bivector language as for vectors:
\begin{equation}
u_m = \frac{1}{2\mu_0} |\bivec{b}|^2
\quad.
\end{equation}
But the Poynting vector that represents the flow of energy in an electromagnetic field (and equivalently, the field's momentum) requires modification:
\begin{equation}
\vec{S} = \frac{1}{\mu_0} \bivec{b} \cdot \vec{E}
\quad.
\end{equation}
This has the same form as Eq.~\eqref{eq:magForceMag} ($\vec{F}_m = q\, \bivec{b} \cdot \vec{v}$), so we can base our intuition on Figure~\ref{fig:ChargeForce3D}.
Electromagnetic fields transport energy only when the electric field has a component in the bivector plane. When that is true, the flow of energy is also in the bivector plane, perpendicular to the electric field.

Naturally, in relativistic language there is nothing new to say at all: relativistic treatments of electromagnetic energy and momentum are already formulated in terms of $\fourvec{F}^{\mu\nu}$. The key result is that the stress-energy-momentum tensor is given by $T^{\mu\nu} = \tfrac{1}{\mu_0} \left( \fourvec{F}^{\mu\rho} \fourvec{F}^{\nu}_{\phantom{\nu}\rho} - \tfrac{1}{4} \eta^{\mu\nu} \fourvec{F}^{\rho\lambda} \fourvec{F}_{\rho\lambda} \right)$.

\subsection{Extra dimensions}
\label{sec:ExtraDims}

As we have seen, the traditional pseudovector approach to magnetism is only defined in three dimensions (where a right-hand rule is meaningful), while the bivector formalism remains natural in two-dimensional examples and in four-dimensional space-time. In fact, if we imagine a world with $d$ dimensions of space (as inspired for example by string theory), the bivector description and equations that we have seen continue to provide a meaningful and intuitive description of magnetic phenomena.

Taking 4-dimensional space as an example, the electric field is described by a 4-component vector $\vec{E} = (E_x, E_y, E_z, E_w)$. But as a bivector, the magnetic field requires a $4\times 4$ antisymmetric matrix:
\begin{equation}
\label{eq:bivecComponents4D}
\bivec{b} = 
 \begin{pmatrix}
0 & b_{xy} & b_{xz} & b_{xw} \\
-b_{xy} & 0 & b_{yz} & b_{yw} \\
-b_{xz} & -b_{yz} & 0 & b_{zw} \\
-b_{xw} & -b_{yw} & -b_{zw} & 0
\end{pmatrix} \quad.
\end{equation}
Because of the antisymmetry, this has  $\binom{d}{2} = \frac{d(d-1)}{2} = 6$ independent components: too many to map into a 4-dimensional vector.

Understanding bivectors geometrically in four or more dimensions requires some care.
In $d$ dimensions, it is possible to represent any bivector as a sum of not more than $d/2$ tiles (rounded down), each corresponding to a wedge product of two vectors.
(A proof is sketched in Appendix~C of \cite{Jensen:2022rot}.) 
This means that every bivector in three dimensions can be represented as a single tile (for example, a sum like $\hat{x} \wedge \hat{y} + \hat{y} \wedge \hat{z}$ can be written as $(\hat{x}-\hat{z})\wedge\hat{y}$), but in four dimensions most require two tiles (e.g.\ $\hat{x}\wedge\hat{y} + \hat{z}\wedge\hat{w}$ cannot be simplified). 
If all the dimensions are spacelike, we can even require that all of the vectors in those wedge products be mutually orthogonal, so the tiles are fully orthogonal as well.

As an explicit example, consider a point particle of charge $q=\qty{1}{\micro\coulomb}$ with velocity $\vec{v} = (9,0,20,-12)\, \unit{\meter\per\second}$ moving in a uniform magnetic field $\bivec{b} = [4 \hat{x}\wedge(\hat{y}+\hat{z}) - 3 (\hat{z}-\hat{y})\wedge\hat{w}]\,\unit{\milli\tesla}$. (All four vectors in these two wedge products happen to be orthogonal.) Then according to Eq.~\eqref{eq:magForce}, the force on the particle is
\begin{equation}
\label{eq:magForce4D}
\vec{F} = 
 \begin{pmatrix}
0 & 4 & 4 & 0 \\
-4 & 0 & 0 & 3 \\
-4 & 0 & 0 & -3 \\
0 & -3 & 3 & 0
\end{pmatrix}
 \begin{pmatrix}
9 \\ 0 \\ 20 \\ -12
\end{pmatrix}\ \unit{\nano\newton}
 =
 \begin{pmatrix}
80 \\ -72 \\ 0 \\ 60
\end{pmatrix}\ \unit{\nano\newton}
\quad.
\end{equation}
This resulting force is perpendicular to the velocity, as usual: the particle will change direction but not speed, and we can expect that it will follow a path that traces simultaneous circles (at different rates) in the planes of the two tiles that constitute $\bivec{b}$.

This may be easier to visualize if we first split the velocity into two pieces, each orthogonal to one of the two planes: $\vec{v} = \vec{v}_1 + \vec{v}_2 = (9,10,10,0)\, \unit{\meter\per\second} + (0,-10,10,-12)\, \unit{\meter\per\second}$. Then $\vec{v}_1$ contributes a force of $\vec{F}_1 = (80,-36,-36,0)\, \unit{\nano\newton}$ (in the plane of the first tile, as visualized in Fig.~\ref{fig:ChargeForce3D}), and $\vec{v}_2$ contributes a force of $\vec{F}_2 = (0,-36,36,60)\, \unit{\nano\newton}$ (in the plane of the second tile). If we group the two resulting accelerations with their corresponding velocities, each piece of the overall velocity will remain in its corresponding plane as the system evolves in time.

Extending these results to $(d+1)$-dimensional spacetime follows exactly the pattern established in Sec.~\ref{sec:Relativity}. The electric and magnetic fields combine into a $(d+1)$-dimensional bivector. Its components form a $(d+1)\times(d+1)$ antisymmetric matrix whose top row (after $b^{tt}=0$) corresponds to the electric field.

\subsection{Classification of fields}
\label{sec:FieldClassification}

For a purely spatial magnetic field (whether in three dimensions as usual, or in extra dimensions as in the previous section), fields can be classified by the algebraic \textit{rank} of the bivector field matrix. This is always an even number, as follows from the result quoted in the previous section that a bivector in $d$ dimensions can be represented by a sum of at most $d/2$ two-dimensional tiles. In three dimensional space, that makes field classification very simple: either there is no magnetic field at all, or the matrix rank is two and the field can be represented by a single tile. But higher dimensions allow a wider variety of algebraically distinct field configurations, as shown in the previous example.

In Lorentzian spacetime, the story becomes substantially more complicated, and a full discussion of the algebraic classification of bivectors is beyond the scope of this article. In Section~\ref{sec:Relativity} we made a simple division of the electromagnetic field in four-dimensional spacetime into a ``magnetic'' tile that was purely spatial plus an ``electric'' tile with one edge spatial and the other purely in the time direction.
This simple electric/magnetic division is always possible, but under some circumstances the two pieces of that sum can be simplified into a single tile.

In brief, if $\bivec{\fourvec{F}}$ can be represented by a single tile that contains a timelike four-vector, then there exists a reference frame where the field is entirely electric ($\bivec{\fourvec{F}}$ has no space-space components). If it can be represented by a single tile that contains only spacelike four-vectors, there is a reference frame where the field is entirely magnetic. And if it can be represented by a single tile containing exactly one null direction, then the fields are typical of a plane wave propagating in that null direction. Otherwise, $\bivec{\fourvec{F}}$ can be written as a sum of two mutually orthogonal tiles (one timelike and ``electric'', one spacelike and ``magnetic''). This classification is the geometric reason that some electromagnetic field configurations have a reference frame where they look purely electric or purely magnetic while others don't.

\section{Bivector math reference}
\label{sec:BivectorReference}

We collect here a number of formal results in bivector mathematics, with both algebraic and geometric interpretations.
(A geometric interpretation of bivector addition can be found in Ref.~\cite{Jensen:2022rot}.) 
For clarity, in this reference we will use the symbols $\vUv$, $\vVv$, and $\vWv$ to refer to generic ordinary vectors (``polar vectors''), while the symbols $\pvLv$, $\pvMv$, and $\pvNv$ will refer to generic pseudovectors (``axial vectors'') whose bivector equivalents are $\bvLv$, $\bvMv$, and $\bvNv$, respectively.

Here, we will relate pseudovector product equations to their bivector equivalents. The results are presented in index notation and involve the totally antisymmetric Levi-Civita symbol $\epsilon_{ijk}$, where $\epsilon_{xyz}=+1$, $\epsilon_{yxz}=-1$, etc.
As we will see, the final bivector forms do not involve $\epsilon_{ijk}$ at all. Although the derivations here are entirely in Cartesian three dimensional space for clarity, the final bivector results in index notation generalize directly to curved space in any dimension (where bivectors are antisymmetric rank-2 contravariant tensors).

In most cases, the equivalences can be derived using the identity
\begin{equation}
\label{eq:EpsilonSquared}
\sum_{i=1}^3 \epsilon_{ijk} \epsilon_{i\idxb\idxc} = \delta_{j\idxb} \delta_{k\idxc} - \delta_{j\idxc} \delta_{k\idxb} \quad,
\end{equation}
though at times a more general identity is needed:
\begin{align}
\label{eq:GenEpsilonSquared}
\nonumber
\epsilon_{ijk} \epsilon_{\idxa\idxb\idxc}
 = {}&\delta_{i\idxa} \delta_{j\idxb} \delta_{k\idxc}
  + \delta_{i\idxb} \delta_{j\idxc} \delta_{k\idxa}
  + \delta_{i\idxc} \delta_{j\idxa} \delta_{k\idxb} \\
  & - \delta_{i\idxa} \delta_{j\idxc} \delta_{k\idxb}
  - \delta_{i\idxb} \delta_{j\idxa} \delta_{k\idxc}
  - \delta_{i\idxc} \delta_{j\idxb} \delta_{k\idxa} \;.
\end{align}

For any pseudovector $\pvLv$ (in three dimensions), we can define a unique ``dual'' bivector $\bvLv$ whose plane is normal to $\pvLv$ , whose magnitude (``area'': $|\bvLv|$) is equal to the vector's magnitude (``length'': $|\pvLv|$), and whose orientation relates to the direction of $\pvLv$ by the right-hand rule. The matrix components $\bvLl_{ij}$ that describe this dual bivector are:
\begin{equation}
\label{eq:vec-to-bivec}
\bvLl_{ij} = \sum_{k=1}^3 \epsilon_{ijk} \pvLl_k
 \equiv \epsilon_{ijk} \pvLl_k \quad.
\end{equation}
This second form illustrates the use of Einstein summation notation (used implicitly through the rest of this section) where repeated indices within a single term are assumed to be summed over all coordinates.%
\footnote{For simplicity, we assume spacelike Cartesian orthonormal coordinates: the distinction between covariant and contravariant indices can be ignored.} 
Explicitly in terms of components, this reads
\begin{equation}
\label{eq:bivec-vec-components}
\bvLl_{ij} 
 = \begin{pmatrix}
0 & \bvLl_{xy} & -\bvLl_{zx} \\
-\bvLl_{xy} & 0 & \bvLl_{yz} \\
\bvLl_{zx} & -\bvLl_{yz} & 0 
\end{pmatrix} = \begin{pmatrix}
0 & \pvLl_z & -\pvLl_y \\
-\pvLl_z & 0 & \pvLl_x \\
\pvLl_y & -\pvLl_x & 0 
\end{pmatrix} \quad.
\end{equation}
We can see that $\bvLl_{ji} = -\bvLl_{ij}$, because of the antisymmetry of $\epsilon_{ijk}$. Comparing components between the two matrices makes the geometric significance clear: for example, the bivector component $\bvLl_{xy}$ in the $xy$-plane is perpendicular to the vector component $\pvLl_z$, with the same magnitude.

The relationship also works in the other direction, with a factor of a half to compensate for double counting (since each independent bivector component appears twice in the matrix):
\begin{equation}
\label{eq:vec-bivec-components}
\pvLl_i = \frac{1}{2} 
         \epsilon_{ijk} \bvLl_{jk}
 = \begin{pmatrix} \bvLl_{yz} \\ \bvLl_{zx} \\ \bvLl_{xy} \end{pmatrix}
 \quad.
\end{equation}

\begin{figure}
\centering
\includegraphics[width=0.8\linewidth]{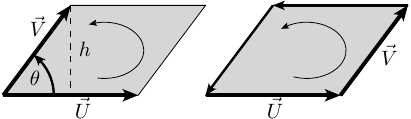}
\caption{\label{fig:WedgeTiles} In the wedge product $\bvLv = \vUv \wedge \vVv$, the two vectors define a parallelogram of area $|\bvLv| = |\vUv| h = |\vUv| |\vVv| \sin\theta$ with a fixed attitude in space. This figure illustrates two ways to visualize its orientation: either as the direction that rotates the first vector toward the second tail-to-tail, or by following the first vector tip-to-tail with the second around the edges.}
\end{figure}

We begin with cross products, since that is where pseudovectors arise. 
The cross product of two ordinary vectors $\pvLv = \vUv \times \vVv$ is expressed in index notation as
\begin{equation}
\pvLl_i = 
 \epsilon_{ijk} \vUl_j \vVl_k \quad.
\end{equation}
(One familiar example of this is the angular momentum of a point particle: $\vec{L} = \vec{r} \times \vec{p}$.)
Using the definition in Eq.~\eqref{eq:vec-to-bivec} (after suitably relabeling summed indices), the symmetries of $\epsilon_{ijk}$, and the identity \eqref{eq:EpsilonSquared}, we find
\begin{align}
\bvLl_{ij} &= 
               \epsilon_{ijk} \pvLl_k
 = 
     \epsilon_{ijk}
    \left( 
            \epsilon_{k\idxb\idxc} \vUl_\idxb \vVl_\idxc \right)
 = 
    \left( 
            \epsilon_{kij} \epsilon_{k\idxb\idxc} \right)
     \vUl_\idxb \vVl_\idxc
\nonumber \\
\label{eq:vec-cross-vec-indices}
&= 
    \left( \delta_{i\idxb} \delta_{j\idxc} - \delta_{i\idxc} \delta_{j\idxb} \right)
    \vUl_\idxb \vVl_\idxc
 = \vUl_i \vVl_j - \vUl_j \vVl_i \quad.
\end{align}
This is the definition of the wedge product:
\begin{equation}
\label{eq:vec-cross-vec-wedge}
\boxed{%
\pvLv = \vUv \times \vVv \qquad \longleftrightarrow \qquad
\bvLv = \vUv \wedge \vVv%
}
\quad.
\end{equation}
This can be visualized as explained in Figure~\ref{fig:WedgeTiles}. Like the cross product, this is antisymmetric: $\vUv \wedge \vVv = -\vVv \wedge \vUv$.

If vectors are expressed in a basis of coordinate unit vectors, $\vUv = \vUl_x \xh + \vUl_y \yh + \vUl_z \zh$, then bivectors will be naturally expressed in a basis of coordinate unit bivectors: $\bvLv = \bvLl_{xy} \xh\wedge\yh + \bvLl_{yz} \yh\wedge\zh + \bvLl_{zx} \zh\wedge\xh$. It's important to recognize that an expression like this should include \emph{either} an $\xh\wedge\yh$ term or a $\yh\wedge\xh$ term but not both: because $\yh\wedge\xh = -\xh\wedge\yh$, including both terms would be redundant. 

\begin{figure}
\centering
\begin{overpic}[percent,width=3.5cm]{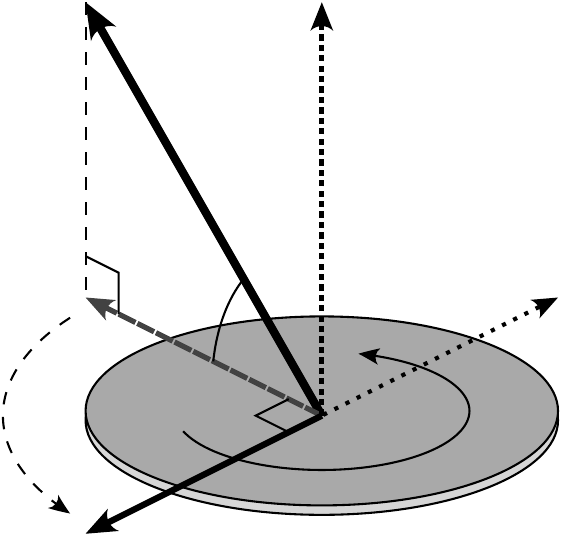}
  \put(67,16){$\bvLv$}
  \put(60,82){$\pvLv$}
  \put(30,79){$\vUv$}
  \put(21,-7){$\vUv \cdot \bvLv$}
  \put(87,45){$\bvLv \cdot \vUv = -\vUv \cdot \bvLv$}
  \put(33,40){$\theta$}
\end{overpic}
\vspace{1mm}
\caption{\label{fig:VecTileMult} Finding the (dot) product of a vector $\vUv$ with a bivector $\bvLv$ is a two-step process. First, the ``dot'' always means to project the vector into the plane of the bivector. Then, order matters: in $\vUv \cdot \bvLv$ the bivector comes \emph{after} the projection, so rotate by \ang{90} in the tile's orientation direction. But in $\bvLv \cdot \vUv$ the bivector comes \emph{before} the projection, so rotate by \ang{90} \emph{opposite} the tile's orientation. Either way, the resulting magnitude is $|\vUv|\,|\bvLv|\,\cos\theta$, where $\theta$ is the angle between the vector and the bivector plane, and the $\cos\theta$ comes from the projection. (This matches the formula for cross product magnitude: if $\pvLv$ is the pseudovector corresponding to $\bvLv$, then its angle from $\vUv$ is $\ang{90}-\theta$ and thus $|\pvLv\times\vUv|=|\pvLv|\,|\vUv|\,\sin(\ang{90} - \theta)=|\pvLv|\,|\vUv|\,\cos\theta$.)}
\end{figure}

Meanwhile, the cross product of  a pseudovector and an ordinary vector is an ordinary vector, $\vVv = \pvLv \times \vUv$, and can be expressed in index notation as
\begin{equation}
\label{eq:pseudo-cross-vec-indices}
\vVl_i
 = 
  \epsilon_{ijk} \pvLl_j \vUl_k = 
    \vUl_k \left(
    \epsilon_{kij} \pvLl_j \right)
  = 
  \vUl_k \bvLl_{ki} \quad.
\end{equation}
(A familiar physical example of this is the magnetic force on a charged particle: $\vec{F} = q \vec{v} \times \vec{B}$, where the magnetic field $\vec{B}$ is a pseudovector.)
We can immediately recognize this as matrix multiplication of a row vector times a square matrix, which we will represent symbolically as a dot product:
\begin{equation}
\label{eq:vec-cross-pseudo-matrix}
\boxed{%
\vVv =  \pvLv \times \vUv \qquad \longleftrightarrow \qquad
\vVv = \vUv \cdot \bvLv%
}
\quad.
\end{equation}
This can be visualized as explained in Figure~\ref{fig:VecTileMult}.%
\footnote{\label{note:ExplainRotation}One way of understanding the process of projection and rotation in Figure~\ref{fig:VecTileMult} begins by expressing the bivector as a wedge product of two vectors: $\protect\bvLv = \vec{F} \wedge \vec{G}$, so $\bvLl_{ij} = F_i G_j - G_i F_j$. Then $\vUv \cdot \protect\bvLv = (\vUl_i F_i) G_j - (\vUl_i G_i) F_j$: any component of $\vUv$ in the direction of $\vec{F}$ gets rotated to point along $\vec{G}$, and any component along $\vec{G}$ gets rotated to point along $-\vec{F}$. This rotates the projection in the same sense that rotates $\vec{F}$ toward $\vec{G}$.} 
 (Note that the order of the two terms has reversed.)
Given the antisymmetry of $\bvLv$, if we reinterpret $\vUv$ as a column vector we can instead write the matrix product in the opposite order: $\vVv = -\bvLv \cdot \vUv$. Performing the matrix multiplication often feels more familiar in this order.

For expressions written in terms of coordinate unit vectors, expanding this product will lead to terms of the form $\xh \cdot (\xh\wedge\yh)$. Using the triple product identity from note~\ref{note:ExplainRotation}, 
$\vUv \cdot (\vVv \wedge \vWv) = (\vUv \cdot \vVv) \vWv - (\vUv \cdot \vWv) \vVv$, we find:
\begin{equation}
\begin{aligned}
\xh \cdot (\xh\wedge\yh) &= \yh
  &  \yh \cdot (\xh\wedge\yh) &= -\xh
  &  \zh \cdot (\yh\wedge\xh) &= 0 \\
(\xh\wedge\yh) \cdot \xh &= -\yh
  & (\xh\wedge\yh) \cdot \yh &= \xh
  & (\yh\wedge\xh) \cdot \zh &= 0 \\
\xh \cdot (\yh\wedge\zh) &= 0
  &  \yh \cdot (\yh\wedge\zh) &= \zh
  &  \zh \cdot (\yh\wedge\zh) &= -\yh \\
 && \vdots
\end{aligned}
\end{equation}
The pattern is easy to remember: the unit vector dot product will ``cancel out'' a matching unit vector in the wedge product, with a minus sign if the wedge product needs to be reversed to bring them side by side.


\textit{(In the remainder of this section, we will omit the derivations of each bivector expression: the methods are similar to those above, but the details can get tedious.)}

\begin{figure}
\centering
\includegraphics[width=\linewidth]{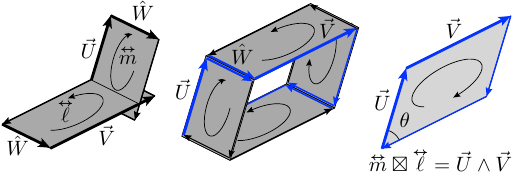}
\caption{\label{fig:TileCommute} The commutator product $\bvMv \boxtimes \bvLv$ of two tiles is non-zero if and only if their planes intersect along a line (with angle $\theta$ between them): geometrically, it is essentially a wedge product of tiles. Reshape the tiles as rectangles with unit length along the shared line ($\hat{\vWl}$), so $\bvMv = \vUv \wedge \hat{\vWl}$ and $\bvLv = \hat{\vWl} \wedge \vVv$.
Then form a ``pipe'' out of two copies of each as shown. The cross section of the pipe (the ``end cap'') is the tile representing the commutator and is equal to $\vUv \wedge \vVv$: a parallelogram of area $|\bvMv|\,|\bvLv| \sin\theta$
whose orientation follows from $\vUv$ through the shared $\hat{\vWl}$ edge to $\vVv$ and back around.}
\end{figure}

Finally, we consider the cross product of two pseudovectors, whose result is another pseudovector: $\pvNv = \pvLv \times \pvMv$. (One example of this is the torque on a magnetic moment in an external magnetic field: $\vec{\tau} = \vec{\mu} \times \vec{B}$.) Converting this to bivector language is messy: it starts like Eq.~\eqref{eq:vec-cross-vec-indices} and uses Eqs.~\eqref{eq:vec-bivec-components} and~\eqref{eq:GenEpsilonSquared}, with the result
\begin{equation}
\bvNl_{ij} =  
  \bvMl_{ik} \bvLl_{kj} - \bvLl_{ik} \bvMl_{kj} \quad.
\end{equation}
The result is recognizable as the matrix product of the two bivectors, antisymmetrized (again in opposite order):
\begin{equation}
\label{eq:pseudo-cross-pseudo-commutator}
\boxed{%
\pvNv = \pvLv \times \pvMv \quad \longleftrightarrow \quad
\bvNv = \bvMv \cdot \bvLv - \bvLv \cdot \bvMv
  \equiv \bvMv \boxtimes \bvLv%
}
\quad.
\end{equation}
We can call this the \defn{commutator product} (or the \defn{bivector cross product}), since this operation is formally called the \defn{commutator} of the two matrices. (The usual notation for a commutator is $[A,B]=AB-BA$.) Its definition implies that it is antisymmetric: $\bvMv \boxtimes \bvLv = -\bvLv \boxtimes \bvMv$. For ``simple'' bivectors that can each be represented by a single tile, this can be visualized as described in Figure~\ref{fig:TileCommute}.

If $\bvMv = \vec{A} \wedge \vec{B}$ and $\bvLv = \vec{U} \wedge \vec{V}$, then
\begin{align*}
\bvMv \boxtimes \bvLv 
 &= (\vec{A} \wedge \vec{V}) (\vec{B} \cdot \vec{U})
    - (\vec{A} \wedge \vec{U}) (\vec{B} \cdot \vec{V}) \\
   &\quad - (\vec{B} \wedge \vec{V}) (\vec{A} \cdot \vec{U})
    + (\vec{B} \wedge \vec{U}) (\vec{A} \cdot \vec{V})
\end{align*}
This allows us to simplify coordinate unit vector expressions like $(\xh\wedge\yh) \boxtimes (\yh\wedge\zh)$:
\begin{equation}
\begin{aligned}
(\xh\wedge\yh) \boxtimes (\yh\wedge\zh) &= \xh \wedge \zh &
  (\xh\wedge\yh) \boxtimes (\zh\wedge\xh) &= \yh \wedge \zh \\
\xh\wedge\yh) \boxtimes (\zh\wedge\yh) &= -\xh \wedge \zh &
  (\xh\wedge\yh) \boxtimes (\xh\wedge\zh) &= -\yh \wedge \zh \\
(\zh\wedge\yh) \boxtimes (\xh\wedge\yh) &= -\zh \wedge \xh &
  (\xh\wedge\yh) \boxtimes (\xh\wedge\yh) &= 0 \\
  & \qquad \qquad \quad \vdots
\end{aligned}
\end{equation}
Again the pattern is straightforward: reverse the wedge products as necessary to put the matching coordinate unit vectors next to each other, multiplying by $-1$ for each reversal, and the result is the wedge product of the remaining two unit vectors.

\bigskip

Having established the formulas for cross products involving pseudovectors, we next consider dot products. The scalar product (or dot product) of two pseudovectors $\alpha = \pvLv \cdot \pvMv$ is a scalar; one example is the energy of a magnetic dipole in an external magnetic field. It can be written as
\begin{equation}
\label{eq:pseudo-dot-pseudo-components}
\alpha = 
  \pvLl_i \pvMl_i = \tfrac{1}{2} 
  \bvLl_{jk} \bvMl_{jk} \quad.
\end{equation}
As in previous equations, the factor of $\frac{1}{2}$ compensates for the duplication of entries in the two antisymmetric matrices. This sum over products of all matrix components is sometimes called the ``double dot product'' and is denoted by a colon:%
\footnote{In this work, the (single) dot product always represents matrix multiplication, or more precisely, tensor index contraction on one index. The double dot product notation used here for contraction on two indices is more common in engineering than in physics: in the traditional physics curriculum, students see second rank tensors only rarely before they learn index notation, at which point these shorthands become unnecessary. In geometric algebra the convention is different: there, roughly speaking, a (single) dot product denotes contraction on as many indices as possible, so the double index contraction of two bivectors would be represented by a single dot product.}
\begin{equation}
\label{eq:pseudo-dot-pseudo-doubledot}
\boxed{%
\alpha = \pvLv \cdot \pvMv \qquad \longleftrightarrow \qquad
\alpha = \tfrac{1}{2} \bvLv \dblcdot \bvMv%
}
\quad.
\end{equation}
As with the vector dot product, we can interpret this scalar product as a measure of whether two bivectors are oriented in the same direction: orthogonal tiles have double dot product zero.
We can use this to define the magnitude of a bivector: $|\bivec{\bvLl}|^2 \equiv \frac{1}{2} \sum_{i,j=1}^3 \bvLl_{ij}^2$.
If we use the antisymmetry of the bivector matrices, we can also interpret this in terms of the trace of the matrix product: $\alpha = -\frac{1}{2} \tr(\bvLv \cdot \bvMv)$.
And if we express $\bvMv = \vUv \wedge \vVv$, then $\alpha = \vUv \cdot \bvLv \cdot \vVv$.

For expressions in terms of coordinate unit bivectors such as $\bvLv = \bvLl_{xy} \xh\wedge\yh + \bvLl_{yz} \yh\wedge\zh + \bvLl_{zx} \zh\wedge\xh$ and similarly for $\bvMv$, this boils down to $\alpha = \bvLl_{xy} \bvMl_{xy} + \bvLl_{yz} \bvMl_{yz} + \bvLl_{zx} \bvMl_{zx}$, exactly like the dot product of two (pseudo)vectors. (By writing the three terms explicitly and only including a single order for each coordinate pair, we have avoided the need for the factor of $\frac{1}{2}$.)

\begin{figure}
\centering
\vspace{3mm}
\begin{overpic}[percent,width=4cm]{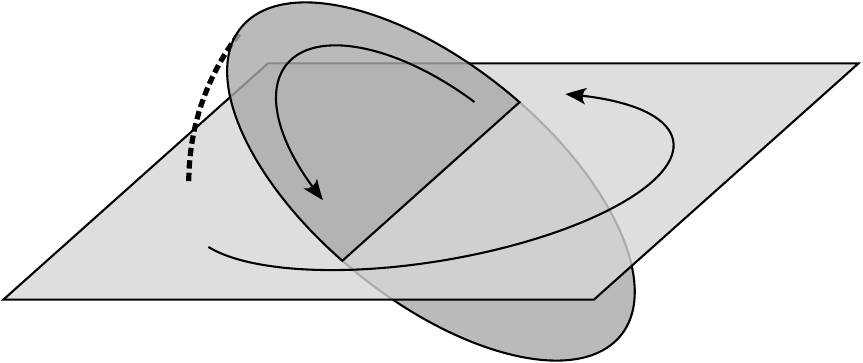}
  \put(40,42){$\bvLv$}
  \put(88,16){$\bvMv$}
  \put(17,29){\large $\theta$}
\end{overpic}
\caption{\label{fig:TileAngle} The scalar product (or double dot product) of two bivectors $\bvLv$ and $\bvMv$ is a product of their same-directed magnitudes. Its value is $\frac{1}{2} \bvLv\dblcdot\bvMv=|\bvLv|\,|\bvMv|\,\cos\theta$.)}
\end{figure}
Geometrically, as shown in Fig.~\ref{fig:TileAngle}, this is a measure of the degree to which the two planes have the same attitude and orientation in space. For two tiles whose planes overlap along a line (which includes any two tiles in three dimensions) the double dot product matches the dot product of their normal vectors: $\frac{1}{2} \bvLv \dblcdot \bvMv = |\bvLv|\, | \bvMv| \cos\theta$, where $\theta$ is the angle between the planes. More generally in higher dimensions (where two planes may only overlap at a point), the double dot product equals zero if \emph{any} vector in one tile is perpendicular to the other tile. The precise geometric meaning in higher dimensions seems complicated: considering the form $\vUv \cdot \bvLv \cdot \vVv$ in light of Figure~\ref{fig:VecTileMult}, we are projecting $\vUv$ into the plane of $\bvLv$, multiplying it by the bivector magnitude, rotating it \ang{90}, and then taking the dot product with $\vVv$.

Finally, the dot product of an ordinary vector and a pseudovector $\phi = \vUv \cdot \pvLv$ is more subtle to understand, because the result $\phi$ is a ``pseudoscalar'' rather than an ordinary scalar:
\begin{align}
\phi &= 
 \vUl_i \pvLl_i
 = \tfrac{1}{2} 
 \epsilon_{ijk} \vUl_i \bvLl_{jk}
 \nonumber \\
 \label{eq:vec-dot-pseudo}
&=  \vUl_x \bvLl_{yz} + \vUl_y \bvLl_{zx} + \vUl_z \bvLl_{xy} \quad.
\end{align}
Every term contains all three spatial coordinates: there are no free indices, just like a scalar quantity. But like a pseudovector, under a reflection like $x \to -x$ all three terms change sign, which a true scalar would not. (An example of this is the electromagnetic invariant $\vec{E} \cdot \vec{B}$.)

\begin{figure}
\centering
\begin{overpic}[percent,width=\linewidth]{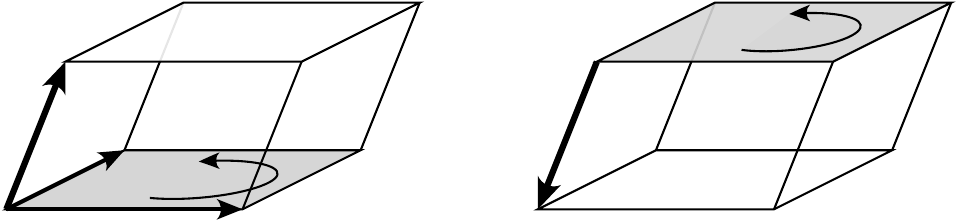}
  \put(16,3.){$\bvLv$}
  \put(0,11){$\vUv$}
  \put(17,-3){$\vVv$}
  \put(7,7){$\vWv$}
  \put(78,18.5){$\bvLv$}
  \put(49,4){$-\vUv$}
\end{overpic}
\caption{\label{fig:Trivector} The wedge product of a vector with a bivector, $\vUv \wedge \bvLv$ (shown at left), can be visualized as an oriented three-dimensional volume formed by extruding the bivector tile along the vector's direction. If the bivector is a wedge product $\bvLv = \vVv \wedge \vWv$, this defines the triple product $\vUv \wedge \vVv \wedge \vWv$. If we choose right-handed coordinates, the component $\phi \equiv \Phi_{xyz}$ is positive if the bivector tile's orientation appears counterclockwise from inside the region (as shown at left) and negative if it appears clockwise from inside (as with $-\vUv \wedge \bvLv$ at right).}
\end{figure}

The appropriate interpretation (which correctly generalizes to higher dimensions) is not a scalar, but the wedge product of $\vUv$ with $\bvLv$:
\begin{subequations}
\label{eq:vec-dot-pseudo-trivec}
\begin{align}
\boxed{%
\phi = \vUv \cdot \pvLv \qquad \longleftrightarrow \qquad
\Phi_{(3)} = \vUv \wedge \bvLv } \\
\label{eq:TrivecIndices}
\Phi_{ijk} = \vUl_i \bvLl_{jk} + \vUl_j \bvLl_{ki} + \vUl_k \bvLl_{ij}%
\quad.
\end{align}
\end{subequations}
(This is symmetric: $\vUv \wedge \bvLv = \bvLv \wedge \vUv$.) 
This is a \defn{trivector}: a totally antisymmetric rank-3 tensor. As shown in Figure~\ref{fig:Trivector}, this can be visualized as an (oriented) region of 3D space, with the bivector tile $\bvLv$ as its base and the vector $\vUv$ showing how it ``extrudes'' into the third dimension. 

If the bivector is a wedge product $ \bvLv = \vVv \wedge \vWv$, this defines the ``triple product'' of (true) vectors: $\vUv \wedge \vVv \wedge \vWv$.
By convention, the pseudoscalar is defined to equal the specific component $\phi \equiv \Phi_{xyz}$ of the trivector: the coefficient of $\hat{x}\wedge\hat{y}\wedge\hat{z}$. (In a right-handed coordinate system, the sign of $\phi$ is positive if the bivector's orientation looks counterclockwise when viewed from inside the parallelepiped.)

Because Equation~\eqref{eq:TrivecIndices} is totally antisymmetric in $i$, $j$, and $k$, it is zero unless the three indices are different. In three dimensions, this means that Eq.~\eqref{eq:vec-dot-pseudo} is the only independent term: that's why it looks like a scalar. But in $d$ dimensions, there are $\binom{d}{3} = \frac{1}{3!} d (d-1)(d-2)$ independent ways of choosing three coordinate labels out of $d$, so that is the number of components. 

\medskip

Replacing pseudovectors with bivectors opens one more possible type of product that is not relevant in three dimensions: the wedge product of two bivectors $\bvLv \wedge \bvMv$. This is a totally antisymmetric rank-4 tensor (a ``quadvector''?), and with four antisymmetrized coordinate indices it is guaranteed to be zero in three dimensions. The explicit coordinate form can be written:
\begin{equation}
\bvLv \wedge \bvMv \to
  \bvLl_{ij} \bvMl_{kl} + \bvLl_{ik} \bvMl_{lj} + \bvLl_{il} \bvMl_{jk} +
  \bvLl_{jk} \bvMl_{il} + \bvLl_{jl} \bvMl_{ki} + \bvLl_{kl} \bvMl_{ij} 
\;.
\end{equation}
This could be imagined as an oriented chunk of 4D space. This does arise in physics: as noted earlier, the electromagnetic invariant $\vec{E} \cdot \vec{B}$ is a pseudoscalar, and in relativistic 4D spacetime a pseudoscalar corresponds to a quadvector. In this case, it arises from the wedge product of the electromagnetic field tensor (a four-bivector) with itself: $\bivec{F} \wedge \bivec{F}$. This zero if $\bivec{F}$ is a ``simple'' bivector (i.e.\ if it can be represented as a single tile).

\pagebreak

\bibliography{bivectors}

\end{document}